\documentclass[11pt,a4paper]{article}
\usepackage{jheppub}
\usepackage{amsmath}
\usepackage[most]{tcolorbox}
\usepackage{dsfont}
\usepackage{ulem}
\usepackage{natbib}
\usepackage{xcolor}
\usepackage[hang,flushmargin]{footmisc}
\usepackage{tikz-cd}
\usepackage{enumitem}
\usepackage{amsfonts,amssymb, amscd,amsmath,latexsym,amsbsy,bm}
\usepackage{stmaryrd}
\usepackage{todonotes}
\usepackage{float}

\usepackage{romannum}

\makeatletter\renewcommand{\@biblabel}[1]{#1.}\makeatother

\newtcolorbox{empheqboxed}{colback=gray!20, 
 colframe=white,
 width=\textwidth,
 sharpish corners,
 top=0mm, 
 bottom=0pt
}

\title{Dynamics for Spin-$1/2$ Particles in Einstein-Gauss-Bonnet Gravity II: Non-Relativistic Case}
\author{$^{1,2}$ E. Maciel}

\affiliation{$^{1}$Unidade Acad\^emica de Engenharia de Produ\c{c}\~ ao, Universidade Federal de Campina Grande,\\
Caixa Postal 10071, 58540-000, Sum\'e, Para\'{\i}ba, Brazil.}

\affiliation{$^{2}$Unidade Acad\^emica de F\'{\i}sica, Universidade Federal de Campina Grande,\\
Caixa Postal 10071, 58429-900, Campina Grande, Para\'{\i}ba, Brazil.}

\emailAdd{eugenio.bastos@professor.ufcg.edu.br}

\abstract{
In this work, I investigate the non-relativistic quantum dynamics of spin-1/2 particles in Einstein-Gauss-Bonnet (EGB) gravity and establish a direct connection between higher-curvature corrections, fermionic dynamics, and the phenomenology of compact objects. Starting from the Dirac Hamiltonian in a static, spherically symmetric EGB spacetime, we perform a Fold-Wouthuysen transformation and derive the effective Hamiltonian, including relativistic kinetic, gravitational, spin-orbit, and higher-curvature contributions. Heisenberg equations are then used to obtain the dynamics of velocity, force, and spin, revealing explicit EGB corrections for both translational motion and spin transport. In particular, the spin-orbit sector induces a modified precession frequency whose fractional deviation from general relativity scales as $\delta_{\Omega}=-4(\xi/M^{2})(M/\rho)^{3}$, providing a clear dimensionless signature of the Gauss-Bonnet coupling. Through Ehrenfest's theorem, we also establish the correspondence between the dynamics of quantum operators and their semiclassical gravitational limit. As an astrophysical application, we consider the stellar-mass black hole A0620-00 and show that prospective relative sensitivities in spin precession on the order of $10^{-3}$ to $10^{-4}$ can probe Gauss-Bonnet couplings in the range of approximately $10^{6}$ to $10^{8}\,{\rm m}^{2}$, depending on the orbital radius. This result identifies fermionic spin precession as a complementary channel for testing gravity with higher-curvature corrections and provides a quantum-mechanical framework connecting modified gravitational dynamics to precision phenomenology in strong-gravity regimes.
}


\begin{document}

\maketitle

\section{Introduction}
\label{sec:introduction}

General relativity (GR) provides a remarkably successful description of gravitational phenomena over a wide range of length and energy scales \cite{carroll2019spacetime,robert1984general,weinberg1973gravitation}. Nevertheless, the occurrence of spacetime singularities and the absence of a complete quantum description of gravity motivate the study of extensions in which higher-curvature terms modify the Einstein-Hilbert action. Among these theories, Gauss-Bonnet gravity occupies a distinguished position. The Gauss--Bonnet invariant arises naturally as the quadratic Lovelock density and leads to second-order field equations in dimensions higher than four \cite{Lovelock:1971einstein,Boulware1985}. Its possible four-dimensional realizations have consequently stimulated extensive investigation, particularly after the proposal of a dimensionally regularized construction of four-dimensional Einstein-Gauss-Bonnet
(EGB) gravity \cite{glavan2020einstein}. Although the original prescription presents conceptual subtleties, several regularized or scalar-tensor formulations have subsequently clarified the dynamical content of related four-dimensional theories\cite{fernandes2020derivation,hennigar2020lower,lu2020horndeski,aoki2020consistent,fernandes2022Review,cvetivc2002black,nojiri2024propagation,nojiri2024,nojiri2017modified}.

Compact objects provide a natural arena in which such higher-curvature effects may become relevant. Black-hole solutions, their stability, quasinormal spectra, shadows, accretion properties, and geodesic structure have therefore been extensively investigated in EGB-inspired models \cite{churilova2021quasinormal,churilova2021quasinormal2, churilova2021wormholes}. At the phenomenological level, Solar-System, binary, gravitational-wave, and compact-object observations have also been employed to constrain the Gauss-Bonnet coupling \cite{clifton2020observational,charmousis2022astrophysical}. These studies show that the strongest deviations from GR are generally expected in regions of large curvature, making stellar-mass compact objects particularly interesting laboratories for searches for higher-curvature gravitational effects.

Most investigations of particle motion in modified gravitational backgrounds are formulated in terms of classical trajectories, geodesics, or wave propagation. A complementary question concerns the dynamics of elementary fermions and, in particular, the way in which their intrinsic spin responds to modifications of the gravitational
field. The Dirac equation in curved spacetime provides the appropriate framework for addressing this problem \cite{brill1957interaction,parker2009quantum,obukhov2013spin}.
For static gravitational backgrounds, the construction of a Hermitian Dirac Hamiltonian and its subsequent Foldy-Wouthuysen (FW) transformation provide a systematic route from relativistic quantum theory to an effective low-energy description \cite{foldy1950dirac,obukhov2001spin,jentschura2013nonrelativistic}. Besides the gravitational modification of the kinetic and potential sectors, this procedure naturally exposes spin-orbit and Darwin-type interactions that are not visible in a purely geodesic description.

The Heisenberg formulation is particularly useful in this context, since it allows gravitational effects to be studied directly through the quantum evolution of physical observables. Position, momentum, and spin operators lead respectively to velocity, force, and spin-precession equations, while the Ehrenfest theorem provides the connection with the corresponding semiclassical dynamics \cite{MACIEL2026170696}. This operator approach therefore offers a unified framework in which both translational and intrinsic spin effects can be identified and their GR and higher-curvature contributions separated. In particular, the spin-orbit interaction provides a natural mechanism through which modifications of the
background geometry can generate observable corrections to the precession of fermionic spin.

In this work, I developed this program for a spin-$1/2$ particle in a static and spherically symmetric EGB background. Starting from the Hermitian Dirac Hamiltonian in isotropic coordinates, we perform the FW transformation and derive the effective nonrelativistic Hamiltonian, retaining the relevant relativistic, gravitational, spin-orbit, Darwin-type, and Gauss-Bonnet contributions. I subsequently employ the Heisenberg equations to derive the velocity and effective force operators and identify the different EGB corrections to the particle dynamics. The resulting force contains not only modified radial and momentum-dependent gravitational contributions but also spin-dependent terms, providing a quantum-mechanical characterization of the departure from the GR dynamics. Through the Ehrenfest theorem, these operator equations are connected with their semiclassical limit.

A central result of our analysis concerns the spin sector. The Gauss-Bonnet correction modifies the gravitational spin-orbit interaction and consequently the spin-precession frequency. The fractional departure from GR exhibits a characteristic cubic radial suppression, $\delta_{\Omega}\propto(\xi/M^{2})(M/\rho)^{3}$, whereas the Gauss-Bonnet contribution to the accumulated precession per orbit decays even more rapidly with distance. These results make explicit that the spin dynamics recovers the GR prediction in the weak-field region while becoming increasingly sensitive to higher-curvature effects closer to the compact object. Spin precession therefore provides a complementary probe of the gravitational background, distinct from the conventional analysis based exclusively on orbitalmotion.

Finally, we connect the quantum-mechanical results with compact-object phenomenology by considering the stellar-mass black-hole system A0620-00, whose mass and comparatively small inferred spin are well constrained observationally \cite{cantrell2010inclination,gou2010spin}. Using its measured mass as an astrophysical benchmark, we translate the EGB correction to the spin-precession frequency into a prospective sensitivity to the Gauss-Bonnet coupling. We find that relative spin-precession sensitivities at the level of $10^{-3}$-$10^{-4}$ can probe coupling scales approximately in the range $10^{6}$-$10^{8}\,{\rm m}^{2}$, depending on the orbital radius. These estimates are not interpreted as present observational bounds, but demonstrate that the fermionic spin dynamics derived from the underlying Dirac theory can be mappedonto quantitatively relevant strong-gravity scales. The present framework thus provides a direct bridge between relativistic quantum dynamics, higher-curvature gravity, and compact-object phenomenology,
and identifies spin transport as an additional channel for testing departures from GR.

The paper is organized as follows.  In Sec.~\ref{sec:EGB}, the formalism is specialized to the EGB geometry and the effective nonrelativistic Hamiltonian is obtained. In Sec.~\ref{sec:dirac}, we review the Dirac theory in a static and spherically symmetric curved spacetime and introduce the FW reduction. The corresponding Heisenberg dynamics, including the velocity, force, and spin evolution, and the spin-precession phenomenology and its application to A0620-00 are discussed inis developed in Sec.~\ref{sec:dynamics}. Finally, our final remarks are presented in Sec.~\ref{sec:conclusions}. Here I used natural units $(\hbar=G=c=1)$.

\section{The Einstein-Gauss-Bonnet Theory}
\label{sec:EGB}
The gravitational framework considered in this work is based on Einstein–Gauss–Bonnet (EGB) gravity, which corresponds to the first nontrivial higher-curvature extension of the Einstein–Hilbert action within the Lovelock construction \cite{Lovelock:1971einstein}. In a $D$-dimensional spacetime, the EGB action can be written as
\begin{equation}
\label{TT1}
S_{\mathrm{EGB}}=\frac{1}{16 \pi} \int d^D x \sqrt{-g}\left[R+\xi \mathcal{G}\right],
\end{equation}
where $R$ is the Ricci scalar, and $\xi$ represents the Gauss–Bonnet coupling parameter\footnote{In the literature, the standard notation for the Gauss-Bonnet parameter is $\alpha$. However, since we will be investigating this theory in the fermionic sector, we will explicitly consider the quantity $\alpha$ to be the Dirac matrix. Therefore, in this work, the Gauss-Bonnet parameter will be described by $\xi$.}. The quantity $\mathcal{G}$ is the Gauss–Bonnet invariant, defined by the particular quadratic combination of curvature tensors
\begin{equation}
\label{TT2}
\mathcal{G}=R_{\mu \nu \rho \sigma} R^{\mu \nu \rho \sigma}
-4R_{\mu \nu}R^{\mu \nu}+R^2.
\end{equation}
An important feature of this combination is that, despite being quadratic in curvature, the corresponding gravitational field equations remain of second differential order in the metric. This property distinguishes the Gauss-Bonnet contribution from generic higher-curvature corrections and constitutes one of the central results of Lovelock’s theorem \cite{Lovelock:1971einstein,zumino1986gravity}.

Spacetime dimensionality plays a fundamental role in determining the dynamical content of the Gauss–Bonnet term. For $D>4$, the variation of $\mathcal{G}$ with respect to the metric yields a non-vanishing contribution to the gravitational field equations. Consequently, EGB gravity provides a genuine modification of Einstein gravity in higher-dimensional spacetimes. However, in the case of ordinary $D=4$ spacetime, this situation changes qualitatively. This occurs because, in general, the integral of the Gauss-Bonnet invariant is proportional, under appropriate topological conditions, to the Euler characteristic of the manifold. Therefore, its variation with respect to the metric contributes only through boundary terms and does not modify the local gravitational equations of motion \cite{Lovelock:1971einstein,zumino1986gravity}. In this sense, the standard Gauss-Bonnet term is dynamically inert in four-dimensional metric gravity. This dimensional obstruction motivated the proposal introduced by Glavan and Lin \cite{glavan2020einstein}. Their construction begins with the EGB theory in a generic dimension $D$ and replaces the original coupling according to $\xi\longrightarrow\xi/{D-4}$. 

The essential idea is that the tensor obtained from the variation of the Gauss–Bonnet sector vanishes when $D=4$. For certain contributions to the equations of motion, this vanishing behavior is proportional to $(D-4)$. The rescaling above is therefore intended to compensate this dimensional factor before the four-dimensional limit is performed. One first derives the equations in arbitrary dimension and only afterwards formally takes the limit $D\rightarrow4$. In this manner, the procedure proposed by Glavan and Lin generates nonvanishing Gauss–Bonnet corrections even though the unrescaled Gauss–Bonnet invariant is topological at exactly $D=4$. It is important, however, to distinguish this limiting prescription from the ordinary four-dimensional variation of the action (\ref{TT1}). If one simply sets $D=4$ from the beginning, no Gauss–Bonnet modification of the local field equations is obtained. The nontrivial contribution arises specifically because the rescaling $\xi/(D-4)$ is introduced before the dimensional limit is taken. Thus, the operations of setting $D=4$ and performing the regularized $D\rightarrow4$ limit are not equivalent. This observation is at the origin of both the phenomenological interest and the theoretical controversy associated with the original four-dimensional EGB proposal.

Indeed, subsequent analyses demonstrated that the Glavan-Lin prescription, considered as a purely metric and generally covariant theory intrinsically defined in four dimensions, is not well defined for completely general geometries \cite{gurses2020there,hennigar2020taking}. In particular, the Gauss–Bonnet tensor cannot in general be factorized globally into $(D-4)$ times a regular four-dimensional covariant tensor. Consequently, the cancellation suggested by the formal rescaling does not define a unique metric theory for arbitrary spacetime configurations. This limitation becomes particularly relevant away from highly symmetric geometries. Nevertheless, the $D\rightarrow4$ construction stimulated the development of several regularized formulations in which the corresponding gravitational dynamics can be given a consistent interpretation. Dimensional reduction and conformal regularization procedures, for example, lead to scalar–tensor theories closely related to the Horndeski class, while alternative Hamiltonian constructions provide another route toward a consistent four-dimensional realization\cite{lu2020horndeski,hennigar2020taking,fernandes2020derivation,aoki2020consistent}. 

These developments indicate that the black-hole geometries originally obtained through the Glavan–Lin prescription can be regarded as effective backgrounds associated with suitable regularized formulations, rather than as solutions of a new purely metric Lovelock theory existing intrinsically in four dimensions. For the purposes of the present investigation, this distinction is particularly important. Our objective is not to establish the Glavan–Lin prescription as a fundamental four-dimensional gravitational theory. Instead, we employ the resulting static and spherically symmetric EGB black-hole geometry as an effective gravitational background and investigate the dynamics of a Dirac particle propagating in this spacetime. Such a viewpoint is especially suitable for probe-field analyses, since the relevant gravitational information entering the fermionic dynamics is encoded directly in the background metric.

Within this effective description, the parameter $\xi$ characterizes deviations from the corresponding general-relativistic geometry. Therefore, the influence of the higher-curvature sector on the fermionic dynamics can be investigated entirely through four-dimensional quantities, without introducing explicit extra-dimensional degrees of freedom in the Dirac sector. In the limit $\xi\rightarrow0$, the general-relativistic background is recovered, whereas non zero values of $\xi$ parameterize the modifications associated with the effective Gauss-Bonnet contribution. This framework provides the gravitational basis for deriving the non relativistic dynamics of spin-$1/2$ particles considered in the following sections.

\subsection{Static Spherically Symmetric Einstein-Gauss-Bonnet Geometry}
We now turn our attention to the spacetime geometry that will provide the gravitational background for the fermionic dynamics investigated in this work. Static and spherically symmetric black-hole solutions constitute one of the most important classes of geometries arising in Einstein–Gauss–Bonnet gravity. In the higher-dimensional formulation, the corresponding line element can be expressed in the Schwarzschild-like form \cite{Boulware1985}
\begin{equation}
\label{TT3}
ds^2=-f(r)dt^2+\frac{dr^2}{f(r)}+r^2 d \Omega_{D-2}^2,
\end{equation}
where $d\Omega_{D-2}^2$ represents the line element of the unit $(D-2)$-dimensional sphere. The entire influence of the gravitational field on this static geometry is therefore contained in the radial function $f(r)$. For the effective four-dimensional EGB black-hole spacetime considered here, the metric function assumes the form \cite{glavan2020einstein,fernandes2020charged}
\begin{equation}
\label{TT4}
f(r)=1+\frac{r^2}{2 \xi}\left[1 \pm \sqrt{1+\frac{8 \xi M}{r^3}}\right].
\end{equation}
Here, $M$ is the gravitational mass parameter and $\xi$ characterizes the strength of the Gauss-Bonnet correction. The Schwarzschild solution is 
recovered when the higher-curvature contribution is removed, provided that the appropriate branch of the solution is selected.

An important property for the equation above is the existence of two algebraically distinct branches, originating from the two possible signs preceding the square root. Their physical properties are substantially different. To identify the branch continuously connected with general relativity, one may consider the limit $\xi\rightarrow0$. Expanding the square root for sufficiently small $\xi$, the solution associated with the minus sign remains finite and approaches the Schwarzschild metric. Conversely, the branch corresponding to the plus sign contains a contribution proportional to $r^2/\xi$ and therefore does not possess a smooth Schwarzschild limit as $\xi\rightarrow0$. The latter branch is also known, in the broader Einstein–Gauss–Bonnet context, to exhibit problematic stability properties \cite{Boulware1985,wheeler1986symmetric}. For these reasons, throughout the present analysis we exclusively adopt the negative branch which is continuously connected to the general-relativistic solution. The departure from Schwarzschild geometry becomes progressively more important as one approaches regions characterized by stronger curvature. This behavior follows directly from the nonlinear radial dependence contained in the square root of Eq. (\ref{TT4}). At sufficiently large radial distances, the Gauss–Bonnet contribution becomes strongly suppressed and the geometry approaches the Schwarzschild spacetime. At shorter distances, however, terms controlled by $\xi$ become increasingly important, modifying the gravitational potential, horizon structure, geodesic motion, tidal effects, and the propagation of test fields. These properties have motivated several investigations of particle motion, shadows, quasinormal modes, tidal forces, and scalar, electromagnetic, and Dirac fields in four-dimensional EGB black-hole backgrounds \cite{guo2020innermost,churilova2021quasinormal,li2021tidal,vieira2023quasibound}.

The higher-curvature contribution nevertheless does not generically eliminate the central pathology of the black-hole geometry. Although the square-root structure modifies the short-distance behavior of the metric function, curvature quantities remain singular as the central region is approached. Therefore, the Gauss–Bonnet correction should not be interpreted, within the present solution, as a mechanism that completely regularizes the black-hole interior. Rather, it changes the manner in which the geometry approaches the high-curvature regime. This distinction is relevant because our analysis concerns a quantum test particle propagating on the gravitational background and does not require extending the fermionic dynamics through the central singularity. The connection with general relativity becomes particularly transparent in the weak-field regime. Assuming that the dimensionless quantity $\xi M/r^3$ is sufficiently small, the square root appearing in Eq. (\ref{TT4}) can be expanded and we find
\begin{equation}
\label{TT5}
f(r) \simeq 1-\frac{2 M}{r}+\frac{4 \xi M^2}{r^4}.
\end{equation}
The first two terms reproduce the familiar Schwarzschild result, whereas the third term represents the leading Gauss-Bonnet correction. Notice that this contribution decreases as $r^{-4}$, in contrast with the Newtonian/Schwarzschild contribution proportional to $r^{-1}$. Consequently, at sufficiently large distances the correction becomes rapidly negligible, while its relative importance increases as the particle probes regions closer to the compact gravitational source. 

This radial hierarchy is precisely what one expects from a higher-curvature modification of the gravitational interaction. In the limit $\xi\rightarrow0$, one immediately recovers
\begin{equation}
f(r)\longrightarrow1-\frac{2M}{r},
\end{equation}
whereas the term $4\xi M^2/r^4$ provides the leading correction associated with the effective higher-curvature sector. The parameter $\xi$ therefore controls the magnitude of the deviation from the Schwarzschild geometry and will subsequently enter the quantum-mechanical operators derived from the Dirac theory. The figure below shows the behavior of the function $f(r)$, Eq. (\ref{TT5}).
\begin{figure}[H]
\centering
\includegraphics[scale=0.5]{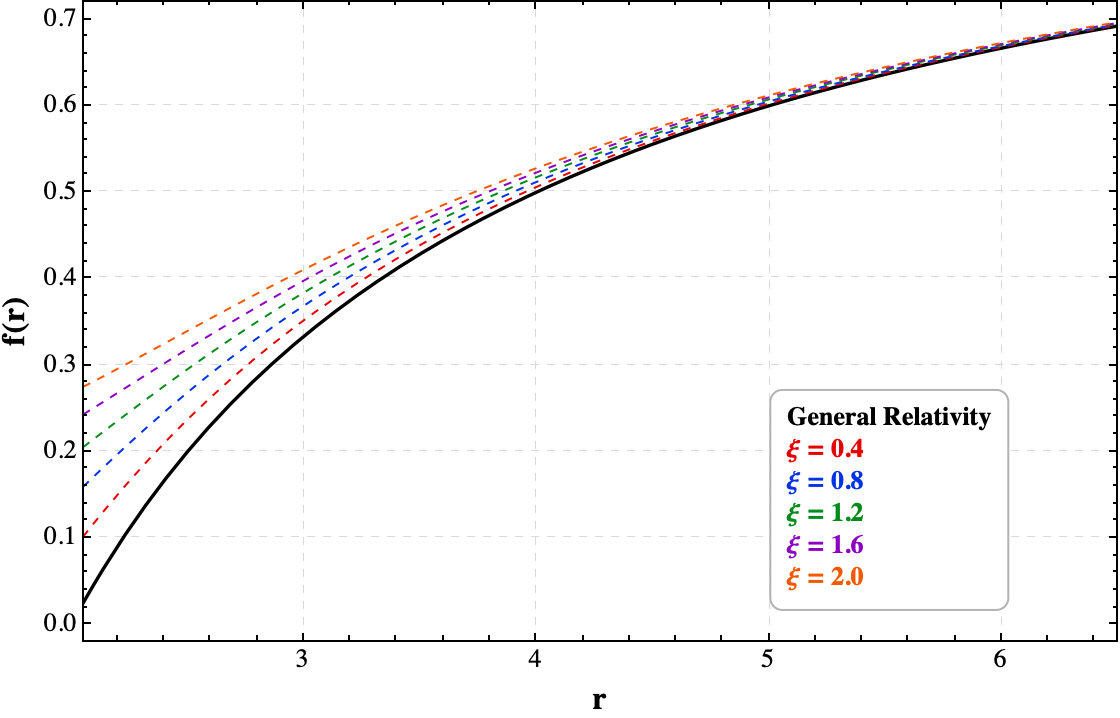}
\caption{\label{fig:metric_function}Radial function $f(r)$, compared with the General Relativity limit for different values of $\xi$.}
\end{figure}
Figure~\ref{fig:metric_function} shows the radial behavior of the metric function $f(r)$ for different values of the Gauss--Bonnet coupling parameter $\xi$, compared with the General Relativity (GR) limit. The EGB corrections become more pronounced in the strong-field region, at smaller radial distances, where increasing $\xi$ produces a systematic deviation from the GR solution. At larger radii, all curves progressively converge toward the GR behavior, indicating that the Gauss-Bonnet contribution is predominantly relevant near the compact object, while the standard gravitational regime is recovered asymptotically.

This behavior becomes especially important when fermionic matter is introduced. The Dirac equation in curved spacetime is sensitive not only to the metric itself but also to the local tetrad fields and the corresponding spin connection. Therefore, modifications of $f(r)$ affect the fermionic dynamics through both the gravitational redshift factors and the derivatives of the metric entering the spinorial connection. In this way, the Gauss–Bonnet contribution is transferred from the classical background geometry to the relativistic Dirac Hamiltonian.
For this purpose, it is convenient to rewrite the geometry (\ref{TT3}) in isotropic coordinates before carrying out the nonrelativistic reduction. In the original Schwarzschild-like coordinates, the radial and angular components of the spatial metric possess different functional structures. By introducing an isotropic radial coordinate, the spatial sector can instead be written in a conformally flat form,

\subsection{Einstein-Gauss-Bonnet Geometry in Isotropic Coordinates.}
For the analysis of the non-relativistic dynamics of a Dirac particle, it is particularly convenient to express the Einstein-Gauss-Bonnet geometry in isotropic coordinates. In this representation, the spatial sector of the metric becomes conformally flat, considerably simplifying the construction of the tetrad fields and the subsequent Foldy–Wouthuysen reduction of the relativistic Hamiltonian. Isotropic coordinates have been widely employed in the study of relativistic quantum particles in static gravitational backgrounds, especially when a separation between relativistic and non-relativistic contributions is required \cite{obukhov2001spin,obukhov2009spin,silenko2008foldy,jentschura2013nonrelativistic,MACIEL2026170696}. Starting from the static, spherically symmetric line element introduced in Eq. (\ref{TT3}), we introduce a new radial coordinate $\rho$ such that the spatial part of the metric takes the conformally flat form
\begin{equation}
\label{ISO1}
ds^2=V^2(\rho)dt^2+W^2(\rho)\left[d\rho^2+\rho^2\left(d\theta^2+\sin^2\theta d\phi^2\right)\right].
\end{equation}
The functions $V(\rho)$ and $W(\rho)$ contain all the information associated with the gravitational background in the isotropic representation. Comparing Eq. (\ref{ISO1}) with the original Schwarzschild-like geometry yields the fundamental relation
\begin{equation}
\label{ISO2}
\frac{d\rho}{\rho} = \frac{dr}{r\sqrt{f(r)}},
\end{equation}
while the angular components imply
\begin{equation}
\label{ISO3}
W^2(\rho)=\frac{r^2(\rho)}{\rho^2}.
\end{equation}
The temporal component is determined directly by
\begin{equation}
\label{ISO4}
V^2(\rho)=f[r(\rho)].
\end{equation}

For the EGB metric considered in the present work, an exact inversion of Eq. (\ref{ISO2}) is not particularly useful for the subsequent quantum-mechanical treatment. Since our goal is to investigate the dominant corrections induced by the Gauss-Bonnet coupling in the weak-field and non-relativistic regimes, we consistently employ the perturbative form of the metric (\ref{ISO1}). The transformation (\ref{ISO2}) can then be solved perturbatively. Up to the order required in the present analysis, one obtains
\begin{equation}
\label{ISO5}
r(\rho)\simeq \rho+M+\frac{M^2}{4\rho} -\frac{\xi M^2}{2\rho^3}.
\end{equation}
Setting $\xi=0$, the first three terms reproduce the weak-field expansion of the standard Schwarzschild transformation
\begin{equation}
r=\rho\left(1+\frac{M}{2\rho}\right)^2.
\end{equation}
Namely the objective of this work is to isolate the dominant modification induced by the Gauss-Bonnet sector within non-relativistic dynamics. To this end, we adopt a perturbative scheme in which the leading contribution from general relativity in the weak-field limit, of order $M/\rho$, is retained alongside the leading term proportional to the Gauss-Bonnet coupling, of order $\xi M^2/\rho^4$. Purely general-relativistic post-Newtonian contributions, of order $M^2/\rho^2$ or higher, are consistently neglected. Consequently, this approximation should not be interpreted as a systematic second-order expansion in the gravitational mass parameter, but rather as a simultaneous expansion in the weak-field limit and the leading order of the EGB coupling. This approach allows for the identification of effects generated specifically by the higher-curvature sector, independently of standard post-Newtonian corrections.

Using Eq. (\ref{ISO5}) the temporal and spatial component of the metric, in corresponding isotropic gravitational function becomes
\begin{equation}
\label{ISO6}
V^2(\rho)\simeq1-\frac{2M}{\rho}+\frac{4\xi M^2}{\rho^4}\quad\mbox{,}\quad W^2(\rho)\simeq1+\frac{2M}{\rho}-\frac{\xi M^2}{\rho^4}.
\end{equation}
Therefore, within the perturbative approximation, the EGB line element can be explicitly written as
\begin{equation}
\label{ISO8}
ds^2\simeq-\left(1-\frac{2M}{\rho}+\frac{4\xi M^2}{\rho^4}\right)dt^2+\left(1+\frac{2M}{\rho}-\frac{\xi M^2}{\rho^4}\right)\Bigg[d\rho^2+\rho^2d\Omega^2\Bigg].
\end{equation}

The Eq. (\ref{ISO8}) makes the structure of the higher-curvature contribution particularly transparent. Schwarzschild terms appear at orders $M/\rho$, whereas the leading EGB modification is proportional to $\xi M^2/\rho^4$. Consequently, the Gauss-Bonnet correction is strongly suppressed in the asymptotic region but becomes progressively more relevant as the particle probes stronger gravitational fields. Furthermore, the opposite signs of the $\xi$-dependent contributions to the temporal and spatial metric components show that the higher-curvature correction modifies the gravitational redshift and the spatial geometry in distinct ways. As an immediate consistency check, taking the limit $\xi\rightarrow0$ yields
\begin{equation}
V^2(\rho)
\longrightarrow 1-\frac{2M}{\rho},
\end{equation}
and
\begin{equation}
W^2(\rho) \longrightarrow 1+\frac{2M}{\rho},
\end{equation}
which corresponds precisely to the weak-field expansion at first order of the Schwarzschild metric written in isotropic coordinates. 

Thus, the general relativity limit is recovered continuously. The isotropic representation (\ref{ISO8}) will be particularly useful in the following analysis. Since the spatial metric is conformally flat, the tetrad structure can be chosen in a considerably simpler way than in the original Schwarzschild-like coordinates. This allows the Dirac Hamiltonian in curved spacetime to be expressed in terms of the two radial functions $V(\rho)$ and $W(\rho)$ and facilitates the identification of its even and odd components with respect to the Dirac matrix $\beta$. The Foldy-Wouthuysen transformation can then be implemented systematically, providing a direct path to the effective non-relativistic Hamiltonian and allowing for the explicit identification of Gauss-Bonnet corrections to the kinetic, gravitational, and spin-dependent sectors. Details of the coordinate transformation leading from Eq. (\ref{TT3}) to Eq. (\ref{ISO8}) are presented in Appendix \ref{app:EGB_isotropic}.

\section{Formalism}
\label{sec:dirac}
The description of spin-$1/2$ particles in a gravitational background requires the extension of the Dirac equation to curved spacetime. Unlike scalar fields, fermionic fields cannot be coupled directly to the spacetime metric alone, since spinors transform according to representations of the local Lorentz group. The appropriate formulation therefore introduces a local orthonormal frame, defined by the tetrad field, which establishes the connection between the curved spacetime manifold and a locally Minkowskian tangent space \cite{davies1980quantum,parker2009quantum,collas2019dirac}. This formalism provides the natural framework for investigating relativistic and nonrelativistic fermionic dynamics in gravitational fields.

In the present work, particular attention is devoted to static and spherically symmetric geometries expressed in spatially isotropic coordinates. This representation is especially convenient for the subsequent Foldy-Wouthuysen (FW) analysis because the spatial part of the metric becomes conformally flat. The corresponding formulation has been extensively investigated in the context of relativistic quantum mechanics in gravitational fields \cite{obukhov2001spin,obukhov2011dirac,silenko2005semiclassical,silenko2008foldy}. At this stage, we keep the gravitational background completely general. The specific metric functions associated with Einstein–Gauss–Bonnet gravity will be introduced only in the following section.
\subsection{The Dirac Equation in Curved Spacetime}

We consider a general, static, and spatially isotropic spacetime, described as the line element Eq. (\ref{ISO1}). In general situations with a geometry of this structure, the functions $V(\rho)$ and $W(\rho)$ are arbitrary functions of the isotropic radial coordinate $\rho$\footnote{At this point, to avoid cluttering the notation, I will express the functions $V(\rho)$ and $W(\rho)$ simply as $V$ and $W$.}. It is also possible to express equivalently in Cartesian isotropic coordinates
\begin{equation}
\label{DI2}
ds^{2}=-V^{2}dt^{2}+W^{2}\delta_{ij}dx^{i}dx^{j}\quad\mbox{,}\quad
\rho=\sqrt{x^{2}+y^{2}+z^{2}}.
\end{equation}
The latter representation makes explicit the conformally flat character of the spatial geometry. The tetrad field $e^{a}_{\mu}$ relates the curved spacetime metric $g_{\mu\nu}$ to the Minkowski metric $\eta_{ab}$ according to
\begin{equation}
\label{DI3}
g_{\mu\nu}=\eta_{ab}e^{a}_{\mu}e^{b}_{\nu}.
\end{equation}
Here, in this work we adopt the signature $\eta_{ab}=\mathrm{diag}(-1,1,1,1)$. Thus, for the metric Eq. (\ref{DI2}) and the definition Eq. (\ref{DI3}) the tetrad field can be chosen as
\begin{equation}
\label{DF2}
e_\mu^a=\left(\begin{array}{cccc}
V & 0 & 0 & 0 \\
0 & W & 0 & 0 \\
0 & 0 & W & 0 \\
0 & 0 & 0 & W
\end{array}\right)\quad\mbox{,}\quad
e_a^\mu=\left(\begin{array}{cccc}
\frac{1}{V} & 0 & 0 & 0 \\
0 & \frac{1}{W} & 0 & 0 \\
0 & 0 & \frac{1}{W} & 0 \\
0 & 0 & 0 & \frac{1}{W}
\end{array}\right).
\end{equation}
The curved-space Dirac matrices are consequently defined through
\begin{equation}
\label{DI6}
\gamma^{\mu}(x)=e_{a}^{\mu}(x)\gamma^{a},
\end{equation}
and satisfy the generalized Clifford algebra
\begin{equation}
\label{DI7}
\{\gamma^{\mu}(x),\gamma^{\nu}(x)\}=2g^{\mu\nu}(x).
\end{equation}
Namely, the Dirac matrices are explicitly written as
\begin{equation}
\label{DF3}
\gamma^{0}=\frac{1}{V}\gamma^0\quad\mbox{,}\quad
\gamma^{i}=\frac{1}{W}\delta_{ij}\gamma^{j}.
\end{equation}

Since the local Lorentz transformations acting on spinors depend on the spacetime point, the ordinary derivative of a spinor does not transform covariantly. One therefore introduces the spinorial covariant derivative
\begin{equation}
\label{DI9}
D_{\mu}\rightarrow D_{\mu}=\partial_{\mu}+\Gamma_{\mu},
\end{equation}
where $\Gamma_{\mu}$ denotes the spin connection. In terms of the tetrad and the Lorentz connection $\omega_{\mu}^{\ ab}$, it can be written as
\begin{equation}
\label{DI10}
\Gamma_{\mu}=\frac{1}{8}
\omega_{\mu}^{\ ab}
[\gamma_{a},\gamma_{b}],
\end{equation}
with
\begin{equation}
\label{DI11}
\omega_\mu^{a b}=e^{a \nu}\left(\partial_\mu e_\nu^b-\Gamma_{\mu \nu}^\lambda e_\lambda^b\right).
\end{equation}
Where the quantity $\Gamma_{\mu \nu}^\lambda$ are the Christoffel symbols. Thus, the covariant Dirac equation consequently assumes the standard form
\begin{equation}
\label{DI12}
\Big[i\gamma^{\mu}(x)D_{\mu}+m\Big]\psi=0.
\end{equation}

Although Eq. (\ref{DI12}) provides the manifestly covariant formulation, the Hamiltonian representation requires some care. In a curved background, the scalar product naturally contains the metric-dependent volume element, and the Hamiltonian obtained directly from the covariant Dirac equation is not necessarily Hermitian with respect to the conventional flat-space scalar product. A suitable redefinition of the spinor wave function allows the dynamics to be expressed in terms of a manifestly Hermitian Hamiltonian. See Ref. \cite{maciel2025gravitational}.

For a static spatially isotropic metric of the form (\ref{DI2}), the resulting Hermitian Dirac Hamiltonian (and Hermitian) can be cast into the remarkably compact form
\begin{equation}
\label{DI13}
\mathcal{H}_{D}=\beta m V+\frac{1}{2}\{\mathcal{F},\boldsymbol{\alpha}\cdot\boldsymbol{p}\},
\end{equation}
where
\begin{equation}
\label{DI14}
\mathcal{F}
\equiv
\frac{V}{W},
\qquad
\boldsymbol{p}=-i\boldsymbol{\nabla},
\end{equation}
and $\{,\}$ denotes the anticommutator. We use the standard definitions
\begin{equation}
\beta=\gamma^{0},
\qquad
\alpha^{i}=\gamma^{0}\gamma^{i}.
\end{equation}
where
\begin{equation}\label{SSS4}
\boldsymbol{\alpha}_{i}=\left(\begin{array}{ll}
0 & \boldsymbol{\sigma}_{i} \\
\boldsymbol{\sigma}_{i} & 0
\end{array}\right),\;\;
\boldsymbol{\Sigma}_{i}=\left(\begin{array}{ll}
\boldsymbol{\sigma}_{i} & 0 \\
0 & \boldsymbol{\sigma}_{i}
\end{array}\right),
\end{equation} 
where $ \boldsymbol\sigma_{i}\; (i=1, 2, 3)$ are the $2\times 2$ Pauli matrices.

The Eq. (\ref{DI13}) is particularly useful because all information concerning the static gravitational geometry is encoded in only two scalar functions, $V$ and $\mathcal{F}=V/W$. The first function modifies the rest-energy sector, whereas $\mathcal{F}$ controls the coupling between the momentum operator and the spatial geometry. In the Minkowski limit,
\begin{equation}
V\rightarrow1,
\qquad
W\rightarrow1,
\qquad
\mathcal{F}\rightarrow1,
\end{equation}
and immediately reduces to the usual form for free particle Dirac Hamiltonian,
\begin{equation}
\mathcal{H}_{D}\rightarrow\mathcal{H}_{D}=\beta m+\boldsymbol{\alpha}\cdot\boldsymbol{p}.
\end{equation}

This formulation is not restricted to a particular gravitational theory. Any static and spherically symmetric geometry that can be expressed in isotropic coordinates can be incorporated by specifying the corresponding functions $V$ and $W$. This property will be essential in the next section, where the Einstein-Gauss-Bonnet geometry will be mapped onto the general structure.
\subsection{The Foldy-Wouthuysen Expansion}
The Foldy–Wouthuysen (FW) transformation provides a systematic procedure for connecting the relativistic Dirac theory with its low-energy, or non relativistic, description \cite{foldy1950dirac,eriksen1958foldy,silenko2008foldy,silenko2015general}. Its main purpose is to construct a representation in which the Hamiltonian is block diagonal with respect to the positive and negative energy sectors. In this representation, the physical interpretation of the position, momentum, and spin operators becomes particularly transparent, while the nonrelativistic expansion can be organized systematically in inverse powers of the particle mass. For this reason, the FW representation is especially useful for identifying gravitational corrections to the kinetic energy, spin-orbit interaction, Darwin term, and other relativistic contributions.

A general Dirac Hamiltonian Eq. (\ref{DI13}), can be written in the form
\begin{equation}
\mathcal{H}_{D}=\beta m+\mathcal{E}+\mathcal{O},
\label{FWgeneral}
\end{equation}
where $\mathcal{E}$ and $\mathcal{O}$ denote the even and odd operators, respectively, defined according to their commutation properties with the Dirac matrix $\beta$,
\begin{equation}
[\beta,\mathcal{E}]=0,
\qquad
\{\beta,\mathcal{O}\}=0.
\label{evenodd}
\end{equation}
The even operators are block diagonal in the standard Dirac representation and therefore do not mix the positive and negative energy components of the spinor. The odd operators, on the other hand, are block off-diagonal and couple these two sectors. The FW transformation consists of a sequence of unitary transformations designed to progressively eliminate the odd part of the Hamiltonian. For a time-independent Hamiltonian, the transformed operator is given by and the resulting Hamiltonian, up to the relevant nonrelativistic orders, takes the familiar form
\begin{equation}
\mathcal{H}_{FW}=\beta m+\mathcal{E}+\frac{\beta}{2m}\mathcal{O}^{2}-\frac{1}{8m^{2}}[\mathcal{O},[\mathcal{O},\mathcal{E}]]+\cdots .
\label{FWexpansion}
\end{equation}
This expression illustrates an important feature of the FW approach: the square of the odd operator generates the leading nonrelativistic kinetic contribution, whereas nested commutators with the even part produce higher-order relativistic and spin-dependent interactions. For more details see Appendix \ref{app:FW_general}.

A priori, the FW transformation was devised for the case of a fermion in an electromagnetic field \cite{bjorken1965relativistic,greiner1990relativistic}, however the same strategy can be extended to fermions propagating in gravitational backgrounds. In particular, static metrics written in spatially isotropic coordinates are simpler in terms of application of (\ref{FWexpansion}). Thus
\begin{equation}
\mathcal{E}=\beta m(V-1),
\qquad
\mathcal{O}=\frac{1}{2}
\{\mathcal{F},\boldsymbol{\alpha}\cdot\boldsymbol{p}\}.
\label{EOgravity}
\end{equation}
Therefore, the first term represents the even gravitational modification of the rest energy, whereas the second term contains the odd momentum-dependent contribution that mixes the upper and lower components of the Dirac spinor. The central quantity entering the leading FW correction is the square of the odd operator,
\begin{equation}
\mathcal{O}^{2}=\frac{1}{4}\{\mathcal{F},\boldsymbol{\alpha}\cdot\boldsymbol{p}\}^{2}.
\label{Osquare1}
\end{equation}
Using
\begin{equation}
\alpha_i\alpha_j = \delta_{ij} + i\epsilon_{ijk}\Sigma_k,
\qquad
[p_i,\mathcal{F}] = -i\partial_i\mathcal{F}.
\label{eq:DiracIdentities}
\end{equation}
the operator (\ref{Osquare1}) generates both spin-independent and spin-dependent gravitational contributions. Schematically, one obtains
\begin{equation}
\mathcal{O}^{2}=\mathcal{F}^{2}\mathbf{p}^{2}+\text{terms involving }\nabla\mathcal{F}+\boldsymbol{\Sigma}\cdot\left(\nabla\mathcal{F}\times\mathbf{p}\right) +\cdots .
\label{OsquareStructure}
\end{equation}
The first contribution determines the gravitational modification of the kinetic energy. The terms containing spatial derivatives of $\mathcal{F}$ generate gradient-dependent quantum corrections, whereas
\begin{equation}
\boldsymbol{\Sigma}\cdot\left(\nabla\mathcal{F}\times\mathbf{p}\right)
\end{equation}
has the characteristic structure of a gravitational spin-orbit interaction. Accordingly, after the FW transformation the Hamiltonian (\ref{FWexpansion}) assumes the generic structure
\begin{equation}
\begin{aligned}
& H_{FW}\simeq m V+\frac{\boldsymbol{p}^2}{2 m}-\frac{\boldsymbol{p}^4}{8 m^3}+\frac{1}{4 m}\left\{\frac{V}{W^2},\boldsymbol{p}^2\right\}+ \\
& +\frac{1}{4 m}\Bigg[2 \boldsymbol{\sigma} \cdot\left(\boldsymbol{\nabla}\left(\frac{V}{W}\right) \times \boldsymbol{p}\right)-\boldsymbol{\sigma} \cdot(\boldsymbol{\nabla} V \times \boldsymbol{p})\Bigg]+\frac{1}{4 m^2} \boldsymbol{\nabla}^2\left(\frac{V}{W}\right) .
\end{aligned}
\label{eq:FWsphericalGeneral}
\end{equation}
where the successive terms encode the gravitational potential energy, the modified kinetic contribution, and higher-order relativistic effects associated with gradients of the metric functions. The explicit algebra leading to this expression is presented in Appendix~\ref{app:FW_general}. 

For positive-energy states, the FW Hamiltonian can finally be projected onto the upper two-component sector by taking $\beta\rightarrow 1$ and $\boldsymbol{\Sigma}\rightarrow\boldsymbol{\sigma}$. The resulting Pauli-type Hamiltonian provides the appropriate starting point for investigating the nonrelativistic dynamics of a spin-$1/2$ particle in the gravitational background. In particular, once the explicit functions $V$ and $W$ associated with the Einstein-Gauss-Bonnet geometry are introduced, Eq.~(\ref{eq:FWsphericalGeneral}) allows the Gauss–Bonnet corrections to be systematically separated from the standard general-relativistic contributions. This procedure makes it possible to identify independently their effects on the gravitational potential, kinetic energy, spin-orbit coupling, and other higher-order quantum terms.

An important property of the Foldy-Wouthuysen Hamiltonian obtained above is its manifest Hermiticity \cite{maciel2025gravitational,obukhov2011dirac,obukhov2009spin}. Since the original Dirac Hamiltonian is Hermitian with respect to the appropriate scalar product, the unitary FW transformation preserves this property order by order in the nonrelativistic expansion. In particular, the position-dependent gravitational functions do not, in general, commute with the momentum operator, $[p_i,f(\rho)]=-i\partial_i f(\rho)$ for the arbitrary function $f(\rho)$. Consequently, terms involving both the gravitational background and powers of the momentum must appear in a symmetrized form. This is explicitly ensured by anticommutators such as
\begin{equation}
\left\{f(\rho),\mathbf{p}^{2}\right\}
=
f(\rho)\mathbf{p}^{2}
+
\mathbf{p}^{2}f(\rho),
\end{equation}
for which
\begin{equation}
\left\{f(\rho),\mathbf{p}^{2}\right\}^{\dagger}
=
\left\{f(\rho),\mathbf{p}^{2}\right\},
\end{equation}
provided that $f(\rho)$ is real. The remaining terms are also Hermitian: the functions of $\rho$ entering the scalar gravitational and Darwin-type contributions are real, while
$\boldsymbol{\sigma}\cdot\mathbf{L}$ is Hermitian and commutes with any radial function $f(\rho)$, since $[\mathbf{L},f(\rho)]=0$. Therefore, within the order retained in the
FW expansion, one has
\begin{equation}
H_{\mathrm{FW}}^{\dagger}=H_{\mathrm{FW}},
\end{equation}
which guarantees unitary time evolution and real expectation values for the physical observables derived from the effective nonrelativistic theory. The symmetrized operator structure is thus not merely a matter of notation, but is essential for preserving the self-adjoint character of the effective Hamiltonian in the position-dependent gravitational background.

\section{Dirac Theory in Einstein-Gauss-Bonnet Universe}
\label{sec:dynamics}

Having established the general formulation of the Dirac equation in a static and spherically symmetric gravitational background, we now specialize the discussion to Einstein-Gauss-Bonnet (EGB) gravity. The purpose of this section is to investigate how the higher-curvature corrections encoded by the Gauss-Bonnet parameter modify the nonrelativistic dynamics of a spin-$1/2$ particle. In particular, the Foldy–Wouthuysen (FW) representation provides a convenient framework in which the different physical contributions to the Hamiltonian can be identified separately, including the gravitational potential, kinetic corrections, momentum-dependent gravitational interactions, spin-orbit coupling, and short-range terms.

\subsection{The Non-Relativistic Hamiltonian}
The FW transformation separates the positive and negative energy sectors of the Dirac theory and provides a systematic expansion of the Hamiltonian in inverse powers of the particle mass. Consequently, the resulting representation establishes a direct connection between the relativistic Dirac theory and its nonrelativistic limit, while retaining the relativistic corrections associated with momentum, spin, and spatial inhomogeneities of the gravitational field \cite{foldy1950dirac,bjorken1965relativistic,greiner1990relativistic,silenko2008foldy}. 

As discussed in the previous sections, the static and spherically symmetric geometry (EGB) in isotropic coordinates is defined for Eq. (\ref{ISO8}). Considering the definitions (\ref{ISO6}) it will be possible to structure a Hamiltonian Eq. (\ref{eq:FWsphericalGeneral}) such that the terms for the Dirac Hamiltonian assumes the form
\begin{equation}
\begin{aligned}
H_{\mathrm{FW}}
={}&
m\left(
1-\frac{M}{\rho}
+\frac{2\xi M^{2}}{\rho^{4}}
\right)
+\frac{\mathbf{p}^{2}}{2m}
-\frac{\mathbf{p}^{4}}{8m^{3}}+\frac{1}{4m}
\left\{
1-\frac{3M}{\rho}
+\frac{3\xi M^{2}}{\rho^{4}},
\mathbf{p}^{2}
\right\} \\
&+\frac{1}{4m}
\left(
\frac{3M}{\rho^{3}}
-\frac{12\xi M^{2}}{\rho^{6}}
\right)
\boldsymbol{\sigma}\cdot\mathbf{L}
+
\frac{1}{4m^{2}}
\left(
\frac{8M^{2}}{\rho^{4}}
+\frac{36\xi M^{2}}{\rho^{6}}
\right).
\end{aligned}
\label{eq:FW_EGB}
\end{equation}
Here $\mathbf{L}=\boldsymbol{\rho}\times\mathbf{p}$ is the orbital angular momentum operator. This expression constitutes the central result of this section. It represents the energy of a particle orbiting an Einstein-Gauss-Bonnet black hole. An important consistency check follows immediately by taking the limit $\xi\rightarrow0$. In this case, all Gauss–Bonnet contributions disappear and the Hamiltonian reduces to the corresponding weak-field result associated with the Schwarzschild geometry. Furthermore, in the simultaneous limit $M\rightarrow0$, the gravitational interactions vanish and the standard FW expansion for a free Dirac particle is recovered. These limits demonstrate explicitly that the EGB corrections appearing in Eq.~(\ref{eq:FW_EGB}) constitute genuine higher-curvature modifications of the conventional Dirac dynamics rather than independent interactions introduced at the particle level.

This Hamiltonian contains all the usual terms well-established in the literature \cite{obukhov2009spin,silenko2008foldy,jentschura2013nonrelativistic,silenko2005semiclassical,obukhov2011dirac,MACIEL2026170696}, in addition to contributions from the parameter $\xi$ that defines the (EGB) gravity. Another important quantity appears explicitly in the first term within the parentheses in (\ref{eq:FW_EGB}). From this, we can define the effective potential.
\begin{equation}
V_{\mathrm{eff}}(\rho)
=
-\frac{mM}{\rho}
+
\frac{2m\xi M^{2}}{\rho^{4}}.
\label{eq:Veff_EGB}
\end{equation}
The first term is the standard Newtonian potential generated by a central mass $M$, whereas the second represents the leading EGB correction within the approximation considered here. A particularly important feature is its radial dependence. While the conventional contribution decreases as $\rho^{-1}$, the Gauss–Bonnet correction behaves as $\rho^{-4}$.
The figure below shows the behavior of the effective potential with respect to the isotropic coordinate $\rho$.
\begin{figure}[H]
\centering
\includegraphics[scale=0.52]{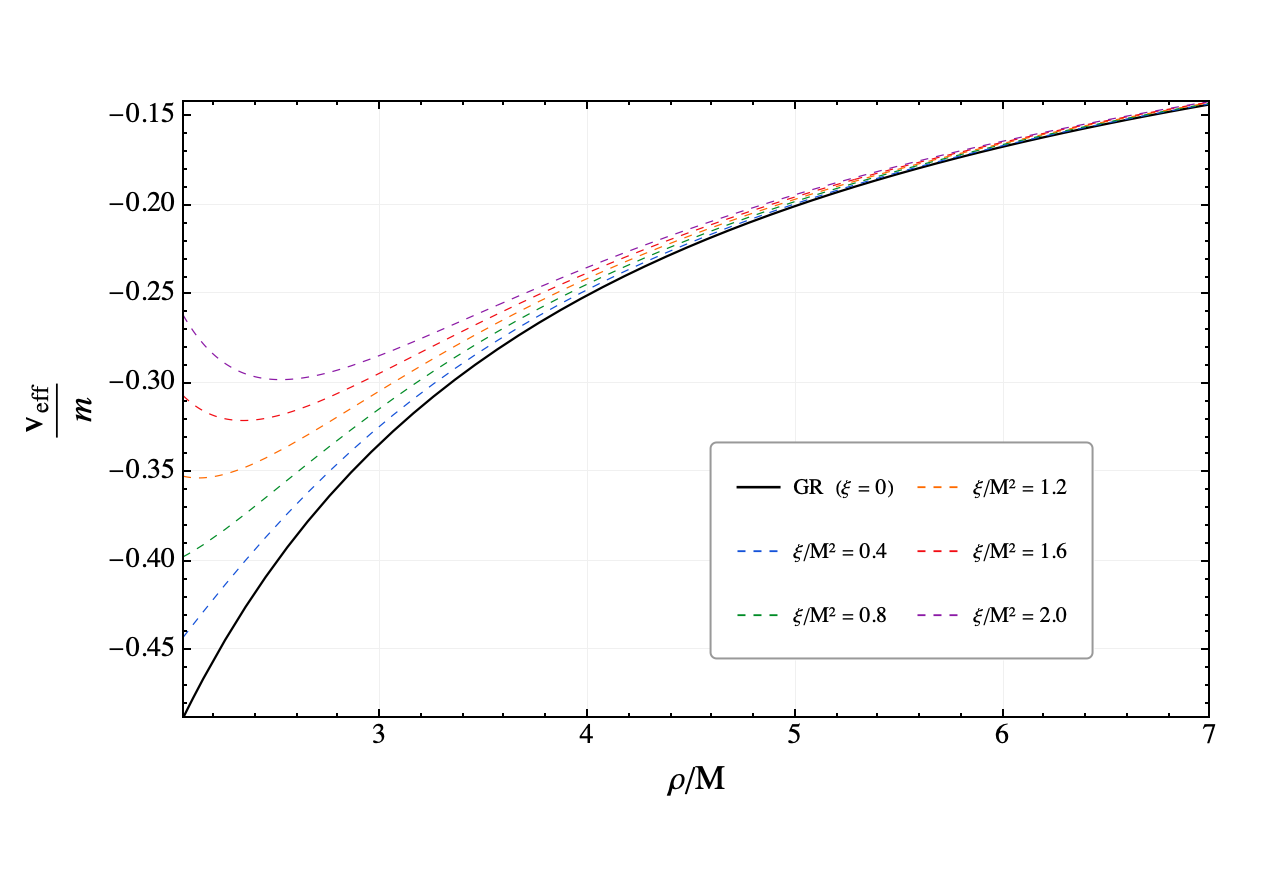}
\caption{\label{effective}Effective gravitational potential $V_{\mathrm{eff}}/m$ as a function of the dimensionless isotropic radial coordinate $\rho/M$.}
\end{figure}

Figure~\ref{effective} shows the radial behavior of the effective gravitational potential experienced by the fermionic particle for different values of the Gauss--Bonnet coupling. The solid black curve corresponds to the general-relativistic (GR) limit, whereas the dashed curves incorporate the Einstein--Gauss--Bonnet (EGB) contribution. A clear departure from the GR behavior emerges at small radial distances, where increasing values of $\xi/M^{2}$ make the effective potential progressively less negative. Thus, for the positive values of the coupling considered here, the EGB correction effectively reduces the depth of the gravitational potential well in the strong-field region. This modification rapidly decreases as $\rho/M$ increases, and the EGB curves progressively converge toward the GR prediction. Such behavior follows directly from the radial dependence of the leading Gauss--Bonnet correction, which scales as $\xi M^{2}/\rho^{4}$ in the weak-field expansion and is therefore suppressed much more rapidly with radial distance than the leading Schwarzschild contribution proportional to $M/\rho$. The figure consequently provides a direct visualization of the regime in which the Gauss--Bonnet sector produces the most significant deviations from GR: the corrections are concentrated in the vicinity of the compact object, while the standard general-relativistic behavior is progressively recovered at larger radial distances \cite{Lovelock:1971einstein,Boulware1985}.

The Eq. (\ref{eq:FW_EGB}) can also be structured into three sectors, namely:
\begin{equation}
\label{FW1}
H_{\text{FW}}=H_{\text{Free}}+H_{\text{Schw}}+H_{\text{EGB}}.
\end{equation}
Where
\begin{equation}
H_{\text {Free }}=m+\frac{\mathbf{p}^2}{2 m}-\frac{\mathbf{p}^4}{8 m^3},
\end{equation}
is the Hamiltonian of the free terms. This Hamiltonian contains the rest energy, the kinetic term, and its correction. The remaining terms are
\begin{equation}
\label{FW2}
H_{\text {Schw }}=-\frac{m M}{\rho}-\frac{3 M}{4 m}\left\{\frac{1}{\rho}, \mathbf{p}^2\right\}+\frac{3 M}{4 m \rho^3} \boldsymbol{\sigma} \cdot \mathbf{L}+\frac{2 M^2}{m^2 \rho^4},
\end{equation}
and
\begin{equation}
\label{FW3}
H_{\mathrm{EGB}}=\frac{2 m \xi M^2}{\rho^4}+\frac{1}{4 m}\left\{\frac{3 \xi M^2}{\rho^4}, \mathbf{p}^2\right\}-\frac{3 \xi M^2}{m \rho^6} \boldsymbol{\sigma} \cdot \mathbf{L}+\frac{9 \xi M^2}{m^2 \rho^6}.
\end{equation}
which represent the Hamiltonian of the standard Schwarzschild gravitational sector and the Hamiltonian emerging from the contribution of the $\xi$ (EGB) parameter, respectively. For ordinary spacetime, the Schwarzschild Hamiltonian (\ref{FW2}) consist of a first term representing the Newtonian gravitational potential, a gravitational kinetic term enclosed in brackets, and a spin-orbit term. The last term is a relativistic correction to the Darwin term.

The Hamiltonian for (EGB) sector Eq.(\ref{FW3}) provides a direct separation of the different physical contributions governing the nonrelativistic dynamics of a Dirac particle in this gravitational background. The first term represents the leading modification induced by the Gauss–Bonnet sector for Newtonian potential. The anticommutator involving $\mathbf{p}^{2}$ describes the coupling between the particle momentum and the spatially varying gravitational field, preserving the Hermiticity of the Hamiltonian and encoding the gravitational modification of the kinetic sector. Particularly relevant for the fermionic dynamics is the term proportional to $\boldsymbol{\sigma}\cdot\mathbf{L}$, which represents the gravitational spin–orbit interaction. Its Schwarzschild contribution, proportional to $M/\rho^{3}$, is supplemented by an EGB correction proportional to $-\xi M^{2}/\rho^{6}$, showing that the higher-curvature effects modify directly the coupling between the intrinsic spin and the orbital motion of the particle. Finally, the last term arise from higher-order contributions in the FW expansion and constitute local quantum-gravitational corrections to the effective energy, arise of Darwin term. Therefore, the EGB parameter $\xi$ affects not only the effective gravitational potential but also the momentum-dependent, spin–orbit, and higher-order quantum sectors of the fermionic Hamiltonian, with corrections characterized by a stronger radial dependence than their corresponding GR contributions. This behavior makes the (EGB) modifications increasingly relevant, relative to the GR terms, as the particle approaches the strong-field region, within the domain of validity of the weak-field and FW expansions.

In particular, one the most relevant term in the present analysis is the spin-orbit term for the (EGB) geometry
\begin{equation}
H_{\mathrm{SO}}^{\mathrm{EGB}}
=
-\frac{3\xi M^{2}}{m\rho^{6}}
\boldsymbol{\sigma}\cdot\mathbf{L}.
\label{eq:spin_orbit_EGB_correction}
\end{equation}
Its $\rho^{-6}$ dependence makes the spin sector especially sensitive to the short-distance structure of the geometry. Indeed, the relative importance of the Gauss–Bonnet correction grows much more rapidly toward the central object than that of the conventional spin–orbit interaction. This result is physically significant because it demonstrates that higher-curvature effects are not restricted to the orbital dynamics: they also modify the coupling between the intrinsic quantum degree of freedom of the fermion and the surrounding spacetime.

From the usual spin-orbit coupling for the Schwarzschild geometry, third term in Eq. (\ref{FW2}) is useful to introduce the radial spin-orbit coefficient
\begin{equation}
C_{\mathrm{SO}}(\rho)
=
\frac{3M}{4m\rho^{3}}
-\frac{3\xi M^{2}}{m\rho^{6}},
\label{eq:C_SO}
\end{equation}
such that
\begin{equation}
\label{FW0}
H_{\mathrm{SO}}
=
C_{\mathrm{SO}}(\rho)
\boldsymbol{\sigma}\cdot\mathbf{L}.
\end{equation}
This form makes explicit the competition between the standard gravitational contribution and the EGB correction. Since they enter with opposite signs for positive $\xi$, the higher-curvature contribution tends to reduce the conventional spin–orbit coupling in the region where it becomes appreciable. This feature provides a direct connection between the EGB parameter and spin-precession effects and will be explored quantitatively below.

In general, the different radial dependencies appearing in Eq.~(\ref{eq:FW_EGB}) provide a useful physical criterion for assessing the relevance of the EGB corrections. At sufficiently large $\rho$, the terms proportional to $\xi$ decay rapidly and the dynamics approaches the conventional general-relativistic regime. Conversely, as the particle moves toward regions of larger curvature, the $\rho^{-4}$ and $\rho^{-6}$ contributions become increasingly important. Therefore, observables associated with orbital motion and, particularly, spin dynamics constitute natural probes of the higher-curvature structure encoded in the EGB geometry. In the following, we analyze these effects on particle dynamics at low energies using the Heisenberg equations of motion, as discussed in a previous work \cite{MACIEL2026170696}.

\subsection{Heisenberg equations of motion}
\label{subsec:heisenberg}

Here, I derive the Heisenberg equations of motion for the position, momentum, and spin operators from the effective Foldy-Wouthuysen Hamiltonian obtained in the previous section. These equations provide the quantum mechanical analogue of the classical geodesic equation and reveal the gravitational spin-orbit coupling responsible for the precession of the spin in the Einstein-Gauss-Bonnet background.

In the Heisenberg picture, the time evolution of an operator \(\mathcal{A}\) with no explicit time dependence is governed by
\begin{equation}
\frac{d\mathcal{A}}{dt} = i\left[H_{\mathrm{FW}}, \mathcal{A}\right],
\label{D1}
\end{equation}
Starting from the Foldy-Wouthuysen Hamiltonian (\ref{eq:FW_EGB}). The velocity operator is defined as the time derivative of the position operator:
\begin{equation}
\mathbf{v} \equiv \frac{d\boldsymbol{\rho}}{dt} = i\left[H_{\mathrm{FW}}, \boldsymbol{\rho}\right].
\label{D3}
\end{equation}
Using the canonical commutation relations
\begin{equation}
[\rho_i, p_j] = i\delta_{ij}, \qquad [p_i, p_j] = 0,
\label{D4}
\end{equation}
we evaluate each contribution to Eq.~(\ref{D3}) separately.
\paragraph{Free-particle contributions.} The standard kinetic terms give
\begin{equation}
i\left[\frac{\mathbf{p}^{2}}{2m}, \boldsymbol{\rho}\right] = \frac{\mathbf{p}}{m},
\label{D5}
\end{equation}
and
\begin{equation}
i\left[-\frac{\mathbf{p}^{4}}{8m^{3}}, \boldsymbol{\rho}\right] = -\frac{\mathbf{p}^{2}\mathbf{p}}{2m^{3}}.
\label{D6}
\end{equation}

\paragraph{Cinetic gravitational term.} For the anticommutator contribution,
\begin{equation}
i\left[\frac{1}{4m}\left\{-\frac{3M}{\rho} + \frac{3\xi M^{2}}{\rho^{4}},\, \mathbf{p}^{2}\right\}, \boldsymbol{\rho}\right]
= \frac{1}{2m}\left\{-\frac{3M}{\rho} + \frac{3\xi M^{2}}{\rho^{4}},\, \mathbf{p}\right\}.
\label{D7}
\end{equation}
\paragraph{Spin-orbit contribution.} Using the commutator
\begin{equation}
i\left[\boldsymbol{\sigma}\cdot\mathbf{L}, \boldsymbol{\rho}\right] = \boldsymbol{\sigma}\times\boldsymbol{\rho},
\label{D8}
\end{equation}
the spin-orbit term contributes
\begin{equation}
i\left[\left(\frac{3M}{4m\rho^{3}} - \frac{3\xi M^{2}}{m\rho^{6}}\right)\boldsymbol{\sigma}\cdot\mathbf{L}, \boldsymbol{\rho}\right]
= \left(\frac{3M}{4m\rho^{3}} - \frac{3\xi M^{2}}{m\rho^{6}}\right)\boldsymbol{\sigma}\times\boldsymbol{\rho}.
\label{D9}
\end{equation}

Collecting all contributions, we obtain
\begin{equation}
\begin{aligned}
\mathbf{v} ={}& \frac{\boldsymbol{p}}{m} - \frac{\boldsymbol{p}^{2}\boldsymbol{p}}{2m^{3}}
+ \frac{1}{2m}\left\{-\frac{3M}{\rho} + \frac{3\xi M^{2}}{\rho^{4}},\, \boldsymbol{p}\right\} \\
&+ \left(\frac{3M}{4m\rho^{3}} - \frac{3\xi M^{2}}{m\rho^{6}}\right)\boldsymbol{\sigma}\times\boldsymbol{\rho}.
\end{aligned}
\label{D10}
\end{equation}
This result shows that the canonical momentum \(\boldsymbol{p}\) is not simply proportional to the velocity in curved spacetime. The last term, proportional to \(\boldsymbol{\sigma}\times\boldsymbol{\rho}\), is transverse to the position vector and represents a spin-dependent modification of the velocity due to the gravitational spin-orbit interaction. The figure \ref{velocity} below shows a simple illustration of a particle experiencing this velocity.
\begin{figure}[H]
\centering
\includegraphics[scale=0.2]{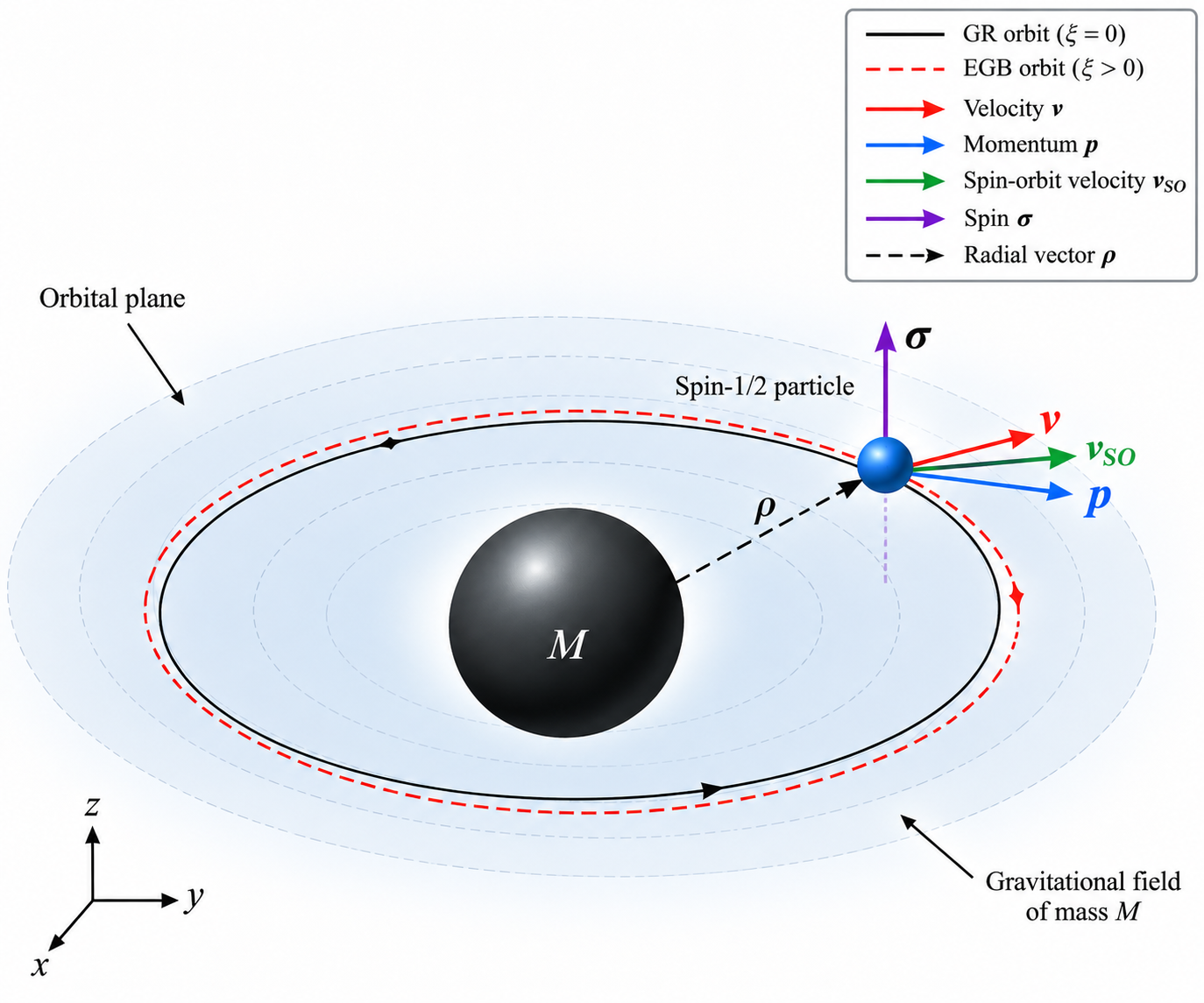}
\caption{\label{velocity}Schematic representation of the orbital cinematic of a spin-$1/2$ particle in the gravitational field of a compact object of mass $M$.}
\end{figure}
The solid black trajectory represents the general-relativistic limit ($\xi=0$), while the dashed red trajectory illustrates the orbital modification induced by the Einstein--Gauss-Bonnet correction $\xi>0$. The radial vector $\boldsymbol{\rho}$ specifies the particle position relative to the central object, whereas $\boldsymbol{p}$ and $\mathbf{v}$ denote its momentum and total velocity, respectively. The vector $\mathbf{v}_{\mathrm{SO}}$ represents the spin--orbit contribution to the velocity operator, and $\boldsymbol{\sigma}$ indicates the spin polarization. The diagram illustrates how the EGB gravitational background, together with the spin--orbit coupling, modifies the fermionic orbital dynamics relative to the GR limit. The relative displacement between the GR and EGB trajectories is shown schematically and is not intended to represent the magnitude of the correction to scale. 

Return to Eq. (\ref{D10}). In particular, the final contribution to the velocity operator-the last term can be identified as a gravitational spin-orbit velocity:
\begin{equation}
\label{FW4}
\mathbf{v}_{\mathrm{SO}} = \left(\frac{3M}{4m\rho^{3}}-\frac{3\xi M^{2}}{m\rho^{6}}\right) \boldsymbol{\sigma}\times\boldsymbol{\rho}.
\end{equation}
This term originates directly from the spin-orbit interaction Eq. (\ref{FW0}) present in the Foldy-Wouthuysen Hamiltonian and, therefore, represents an intrinsic, spin-dependent contribution to the kinematics of the Dirac particle. Since $\mathbf{v}_{\mathrm{SO}}$ is perpendicular to both $\boldsymbol{\sigma}$ and $\boldsymbol{\rho}$, it is oriented along the tangential direction of the orbit when the spin is aligned with the orbital angular momentum. Consequently, the gravitational spin-orbit interaction modifies the relationship between momentum and velocity, such that the latter is no longer simply given by $\boldsymbol{p}/m$. The first term in its coefficient corresponds to the standard General Relativity (GR) contribution, while the second represents the EGB correction. For positive $\xi$, the latter enters with the opposite sign and, therefore, reduces the GR spin-orbit contribution at the perturbative order considered. Furthermore, while the GR spin-orbit velocity varies approximately as $\rho^{-2}$, the EGB contribution varies as $\rho^{-5}$, demonstrating that the modification arising from higher-order curvature is strongly concentrated in the vicinity of the central gravitational source.

\begin{figure}[H]
\centering
\includegraphics[scale=0.5]{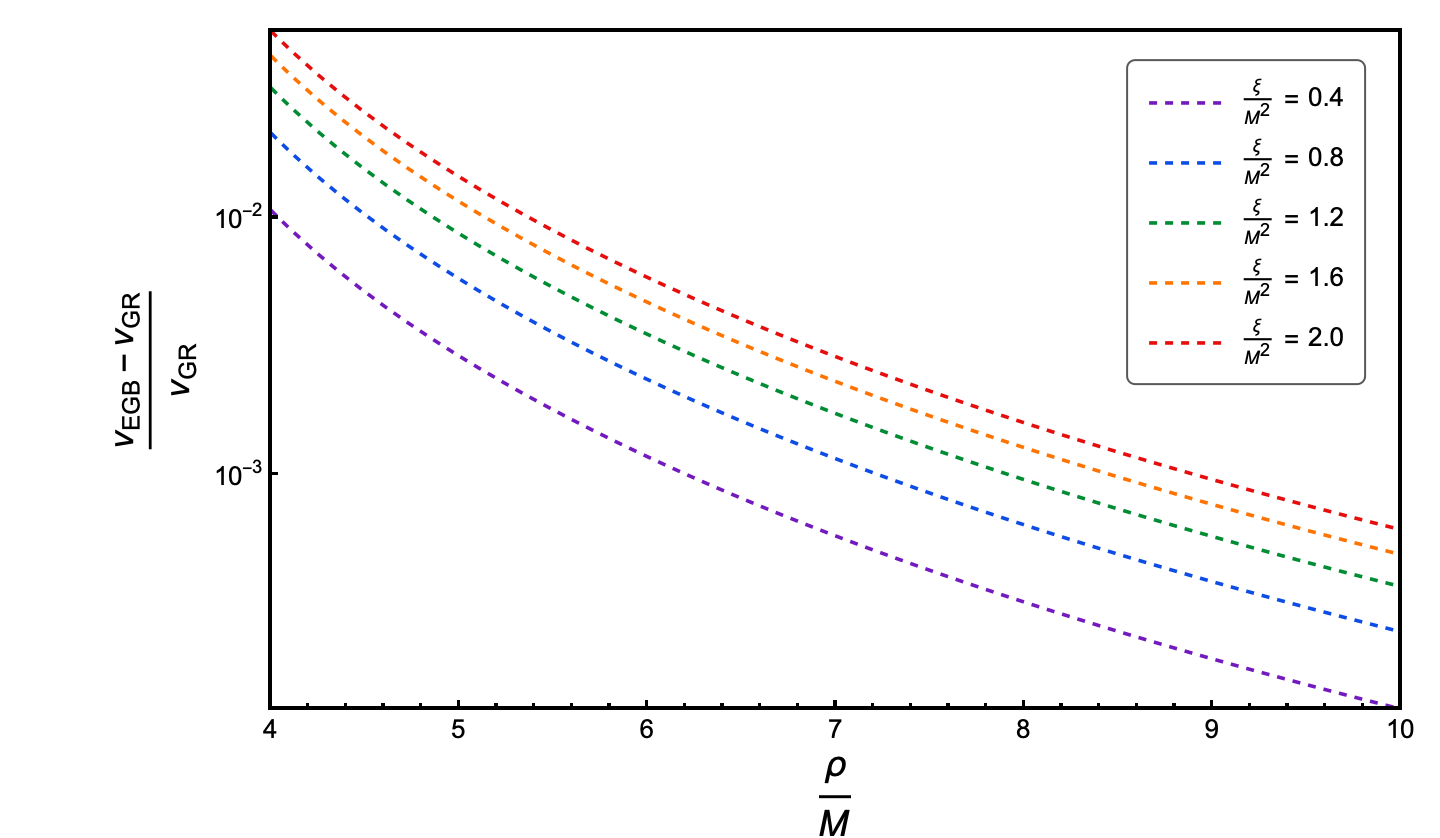}
\caption{\label{fig:velocity2}Relative Einstein-Gauss-Bonnet correction to the velocity of a spin-$1/2$ particle as a function of the dimensionless radial coordinate $\rho/M$.}
\end{figure}

To quantify the influence of Gauss–Bonnet coupling on particle kinematics, we evaluate the relative correction to the tangential velocity, defined as
\begin{equation}
\label{FW5}
\delta_{\bf v}(\rho)=\frac{{\bf v}_{\mathrm{EGB}}-{\bf v}_{\mathrm{GR}}}{{\bf v}_{\mathrm{GR}}},
\end{equation}
for different values of the dimensionless Gauss–Bonnet parameter ($\xi/M^2$). We consider the semi-classical configuration of a circular orbit ($\boldsymbol{p}\perp\boldsymbol{\rho}$), with the spin aligned to the orbital angular momentum ($\boldsymbol{\sigma}\parallel\mathbf{L}$), so that the spin-orbit velocity is in the tangential direction. The dominant order relation for circular orbits ($p/m\simeq\sqrt{M/\rho}$) is then used in conjunction with the semiclassical reduction of the anticommutator in Eq. (\ref{D10}). The resulting behavior is presented in Fig. \ref{fig:velocity2}, where the relative correction is represented as a function of ($\rho/M$), adopting ($mM=1$). The correction is positive for the configuration considered and increases systematically with ($\xi/M^2$), exhibiting, at the same time, a marked radial suppression. In fact, introducing ($x=\rho/M$), ($\bar{\xi}=\xi/M^2$) and ($\mu=mM$), the EGB contribution satisfies
\begin{equation}
\label{FW6}
{\bf v}_{\mathrm{EGB}}-{\bf v}_{\mathrm{GR}}=3\bar{\xi}\left(x^{-9/2}-\mu^{-1}x^{-5}\right).
\end{equation}

The fast decay observed in Fig. \ref{fig:velocity2} therefore arises directly from the higher inverse powers of the radial coordinate associated with the Gauss–Bonnet contribution. The logarithmic scale highlights that the deviation from the GR decreases several orders of magnitude as the radial distance increases, with all curves progressively approaching the GR limit. Furthermore, the approximately proportional separation of the curves reflects the linear dependence on ($\xi$) in the perturbative order considered in the Foldy–Wouthuysen expansion. These results indicate that the modification of the momentum-velocity relationship induced by Gauss–Bonnet coupling is predominantly concentrated at smaller radial distances, while the standard behavior of the RG is quickly recovered as ($\rho/M$) increases.

\subsubsection{Force operator}

The force operator is defined as the time derivative of the canonical momentum:
\begin{equation}
\boldsymbol{F}=\frac{d\boldsymbol{p}}{dt} = i\left[H_{\mathrm{FW}}, \boldsymbol{p}\right].
\label{D11}
\end{equation}
The part of the free Hamiltonian in Eq. (\ref{eq:FW_EGB}) commutes with the operator $\boldsymbol{p}$. Therefore, we will analyze the other terms separately below.

\paragraph{Central potential contributions.} In general, for an arbitrary radial function \(f(\rho)\), we have the fundamental commutator
\begin{equation}
i[f(\rho), \boldsymbol{p}] = -\boldsymbol{\nabla}f(\rho).
\label{D13}
\end{equation}
Applying this to the scalar potential terms in Eq.~(\ref{eq:FW_EGB}), we obtain the central force
\begin{equation}
\boldsymbol{F}_{0} = -\left(\frac{mM}{\rho^{2}} - \frac{8m\xi M^{2}}{\rho^{5}} - \frac{54\xi M^{2}}{m^{2}\rho^{7}}\right)\hat{\boldsymbol{\rho}}.
\label{D14}
\end{equation}

The terms proportional to \(\xi M^2\) in the last parentheses arise from the scalar term \(\frac{1}{4m^{2}}\left(\frac{36\xi M^{2}}{\rho^{6}}\right)\) in Eq.~(\ref{eq:FW_EGB}).
\paragraph{Momentum-dependent gravitational contribution.} For the anticommutator term, we find
\begin{equation}
\mathbf{F}_{p} = -\frac{1}{4m}\left\{\left(\frac{3M}{\rho^{2}} - \frac{12\xi M^{2}}{\rho^{5}}\right)\hat{\boldsymbol{\rho}},\, \boldsymbol{p}^{2}\right\}.
\label{D15}
\end{equation}
This contribution retains the operator ordering required by the quantum nature of the dynamics.
\paragraph{Spin-orbit contribution.} To evaluate the spin-dependent force, we write
\begin{equation}
H_{\mathrm{SO}} = C(\rho)\,\boldsymbol{\sigma}\cdot\mathbf{L},
\label{D16}
\end{equation}
where
\begin{equation}
C(\rho) \equiv \frac{3M}{4m\rho^{3}} - \frac{3\xi M^{2}}{m\rho^{6}}.
\label{D17}
\end{equation}
Using the commutator
\begin{equation}
i[\boldsymbol{\sigma}\cdot\mathbf{L}, \boldsymbol{p}] = \boldsymbol{\sigma}\times\boldsymbol{p},
\label{D18}
\end{equation}
and
\begin{equation}
i[C(\rho), \boldsymbol{p}] = -C'(\rho)\,\hat{\boldsymbol{\rho}},
\label{D19}
\end{equation}
with
\begin{equation}
C'(\rho) = -\frac{9M}{4m\rho^{4}} + \frac{18\xi M^{2}}{m\rho^{7}},
\label{D20}
\end{equation}
we obtain
\begin{equation}
\mathbf{F}_{\mathrm{SO}} = C(\rho)\,\boldsymbol{\sigma}\times\boldsymbol{p} + C'(\rho)\,(\boldsymbol{\sigma}\cdot\mathbf{L})\,\hat{\boldsymbol{\rho}}.
\label{D21}
\end{equation}
Explicitly,
\begin{equation}
\mathbf{F}_{\mathrm{SO}}=\left(\frac{3M}{4m\rho^{3}}-\frac{3\xi M^{2}}{m\rho^{6}}\right)\boldsymbol{\sigma}\times\boldsymbol{p}+\left(\frac{9M}{4m\rho^{4}}-\frac{18\xi M^{2}}{m\rho^{7}}\right)(\boldsymbol{\sigma}\cdot\mathbf{L})\,\hat{\boldsymbol{\rho}}.
\label{D22}
\end{equation}

The first term is transverse to both the spin and momentum, while the second acts in the radial direction and distinguishes different relative orientations of the spin and orbital angular momentum.

\paragraph{Complete force operator.} Combining Eqs.~(\ref{D14}), (\ref{D15}), and (\ref{D22}), the total force operator is
\begin{equation}
\begin{aligned}
\mathbf{F} ={}& -\left(\frac{mM}{\rho^{2}} - \frac{8m\xi M^{2}}{\rho^{5}} - \frac{54\xi M^{2}}{m^{2}\rho^{7}}\right)\hat{\boldsymbol{\rho}}
-\frac{1}{4m}\left\{\left(\frac{3M}{\rho^{2}} - \frac{12\xi M^{2}}{\rho^{5}}\right)\hat{\boldsymbol{\rho}},\, \boldsymbol{p}^{2}\right\} \\
&+ \left(\frac{3M}{4m\rho^{3}} - \frac{3\xi M^{2}}{m\rho^{6}}\right)\boldsymbol{\sigma}\times\boldsymbol{p}
+ \left(\frac{9M}{4m\rho^{4}} - \frac{18\xi M^{2}}{m\rho^{7}}\right)(\boldsymbol{\sigma}\cdot\mathbf{L})\,\hat{\boldsymbol{\rho}}.
\end{aligned}
\label{D23}
\end{equation}

The spin-dependent terms in Eq.~(\ref{D23}) represent a gravitational Stern-Gerlach effect, where the spin orientation influences the trajectory of the particle~\cite{silenko2005semiclassical,obukhov2011dirac}. This is a direct consequence of the coupling between the spin and the gravitational field in the FW Hamiltonian and it constitutes one of the main results of this work.

\begin{figure}[H]
\centering
\includegraphics[scale=0.2]{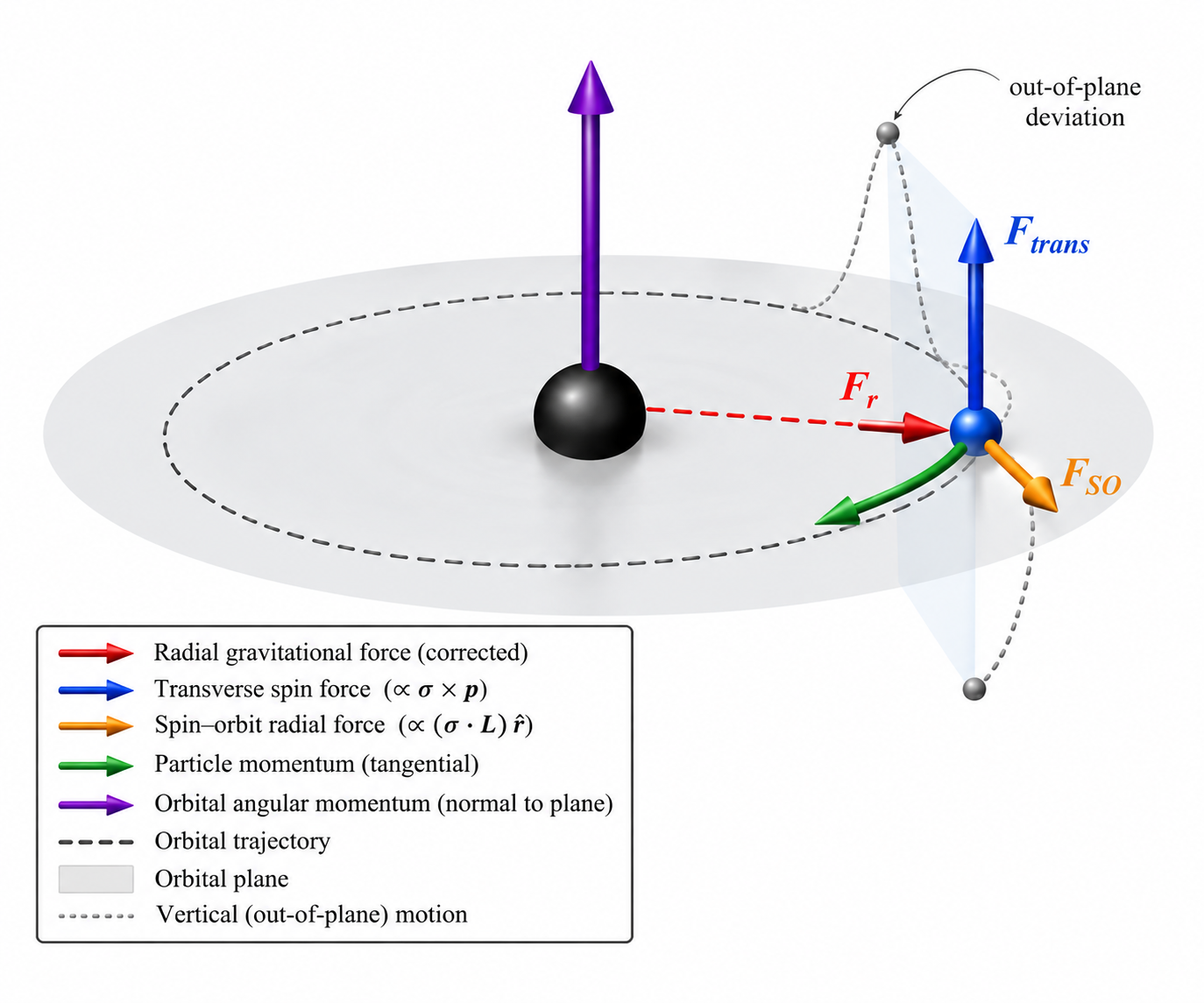}
\caption{\label{fig:force}Schematic representation of the effective force acting on a spin-$1/2$ particle in the gravitational field of an Einstein--Gauss--Bonnet compact object.}
\end{figure}

The equation (\ref{D23}) provides a direct physical characterization of the non-relativistic dynamics of a spin-$1/2$ particle in the Einstein-Gauss-Bonnet gravitational background. The first term represents the central gravitational force. Its main contribution, $-mM/\rho^{2}$ reproduces the Newtonian attractive force, while the proportional terms $\xi M^{2}/\rho^{5}$ and $\xi M^{2}/(m^{2}\rho^{7})$ describe short-range corrections induced by the Gauss-Bonnet sector and higher-order relativistic contributions. Their faster radial decay shows that these effects become increasingly relevant in the strong field region, while they quickly disappear at large distances, where the usual gravitational dynamics are recovered. The second contribution, involving the anticommutator with $\boldsymbol{p}^{2}$, is a momentum-dependent gravitational force originating from the position dependence of the kinetic sector of the Foldy–Wouthuysen Hamiltonian. It therefore represents the coupling between the kinetic movement of particles and the curved geometry of space-time and has no analogue in a purely Newtonian description.The figure (\ref{fig:force}) illustrates in a simple way the dynamics of a particle in an EGB geometry.

The corrected radial gravitational force, $\mathbf{F}_{r}$, acts along the radial direction and contains the conventional gravitational contribution together with the EGB corrections. The spin-dependent sector gives rise to two additional contributions: the transverse spin force, $\mathbf{F}_{\mathrm{trans}}\propto\boldsymbol{\sigma}\times\boldsymbol{p}$, and the spin--orbit radial force, $\mathbf{F}_{\mathrm{SO}}\propto(\boldsymbol{\sigma}\cdot\mathbf{L})\hat{\boldsymbol\rho}$. While the latter modifies the effective radial dynamics according to the relative orientation between the spin and orbital angular momentum, the transverse term generates a force perpendicular to both $\boldsymbol{\sigma}$ and $\boldsymbol{p}$ and may therefore induce an out of plane component of the particle motion. The orbital angular momentum is shown normal to the reference orbital plane, whereas the particle momentum is tangential to the orbit. The trajectories and vector magnitudes are schematic and are not drawn to scale.

Return to Eq. (\ref{D23}). The last two terms explicitly reveal the spin-dependent structure of the gravitational force. The proportional contribution to $\boldsymbol{\sigma}\times\boldsymbol{p}$ is transverse to both the spin polarization and the particle momentum and can consequently be interpreted as a transverse spin-orbit gravitational force, a case analogous to that of the electromagnetic field in \cite{shen2005spin}. It is the strength-level manifestation of the spin-orbit interaction generated by the curved bottom, with the contribution of the EGB modifying its radial dependence through the additional term $\rho^{-6}$. On the other hand, the term proportional to $(\boldsymbol{\sigma}\cdot\mathbf{L})\hat{\boldsymbol{\rho}}$ is purely radial and depends on the relative orientation between the spin and the orbital angular momentum. Thus, states with different spin-orbit alignments experience different effective radial forces. The coexistence of these spin-dependent transverse and radial contributions shows that fermionic dynamics cannot, in general, be represented solely by motion in a scalar gravitational potential. Instead, the EGB geometry modifies both the central orbital dynamics and the coupling between the internal spin degree of freedom and the orbital motion. In the limit $\xi\rightarrow0$, the Gauss-Bonnet corrections disappear and the corresponding general relativistic Foldy-Wouthuysen force is recovered, while in general $\rho$ the main Newtonian contribution dominates.

\subsubsection{Spin dynamics}

The physical spin operator in the FW representation is given by
\begin{equation}
\mathbf{S} \equiv \frac{\boldsymbol{\sigma}}{2},
\label{D24}
\end{equation}
whose equation of motion is
\begin{equation}
\frac{d\mathbf{S}}{dt} = i[H_{\mathrm{FW}}, \mathbf{S}].
\label{D25}
\end{equation}

Since all spin-independent terms commute with \(\mathbf{S}\), only the spin-orbit Hamiltonian \(H_{\mathrm{SO}}\) contributes. Writing
\begin{equation}
H_{\mathrm{SO}} = 2C(\rho)\,\mathbf{S}\cdot\mathbf{L},
\label{D26}
\end{equation}
and using the spin algebra
\begin{equation}
[S_i, S_j] = i\epsilon_{ijk}S_k, \qquad [L_i, S_j] = 0,
\label{D27}
\end{equation}
we obtain
\begin{equation}
\frac{d\mathbf{S}}{dt} = 2C(\rho)\,\mathbf{L}\times\mathbf{S}.
\label{D28}
\end{equation}

Substituting Eq.~(\ref{D17}), we find
\begin{equation}
\frac{d\mathbf{S}}{dt} = \left(\frac{3M}{2m\rho^{3}} - \frac{6\xi M^{2}}{m\rho^{6}}\right)\mathbf{L}\times\mathbf{S}.
\label{D29}
\end{equation}
It is convenient to introduce the spin-precession vector \(\boldsymbol{\Omega}_{\mathrm{EGB}}\) through
\begin{equation}
\frac{d\mathbf{S}}{dt} \equiv \boldsymbol{\Omega}_{\mathrm{EGB}}\times\mathbf{S}.
\label{D30}
\end{equation}
with
\begin{equation}
\boldsymbol{\Omega}_{\mathrm{EGB}} = \left(\frac{3M}{2m\rho^{3}} - \frac{6\xi M^{2}}{m\rho^{6}}\right)\mathbf{L}.
\label{D31}
\end{equation}

The precession vector can be separated into its general relativistic (GR) and EGB contributions:
\begin{equation}
\boldsymbol{\Omega}_{\mathrm{EGB}} = \boldsymbol{\Omega}_{\mathrm{GR}} + \delta\boldsymbol{\Omega}_{\mathrm{GB}},
\label{D33}
\end{equation}
where
\begin{equation}
\boldsymbol{\Omega}_{\mathrm{GR}} = -\frac{3M}{2m\rho^{3}}\mathbf{L}\quad\mbox{,}\quad
\delta\boldsymbol{\Omega}_{\mathrm{GB}} = \frac{6\xi M^{2}}{m\rho^{6}}\mathbf{L}.
\label{D34}
\end{equation}

Equivalently, we may write
\begin{equation}
\boldsymbol{\Omega}_{\mathrm{EGB}} = \boldsymbol{\Omega}_{\mathrm{GR}}\left(1 - \frac{4\xi M}{\rho^{3}}\right),
\label{D36}
\end{equation}
which implies the relative correction
\begin{equation}
\frac{\Omega_{\mathrm{EGB}} - \Omega_{\mathrm{GR}}}{\Omega_{\mathrm{GR}}} = -\frac{4\xi M}{\rho^{3}}.
\label{D37}
\end{equation}
Equation~(\ref{D37}) provides a particularly useful measure of the influence of Gauss-Bonnet coupling on spin dynamics, as it quantifies the fractional deviation of the spin precession frequency in the EGB model from its value in general relativity. By introducing the dimensionless quantities $\bar{\xi}\equiv\xi/M^{2}$ and $x\equiv\rho/M$, the relative correction can be written as $(\Omega_{\mathrm{EGB}}-\Omega_{\mathrm{GR}})/\Omega_{\mathrm{GR}}=-4\bar{\xi}/x^{3}$. This form makes it explicit that the magnitude of the modification is governed by both the dimensionless EGB coupling and the orbital compactness. In particular, for positive $\xi$, the correction is negative, implying $\Omega_{\mathrm{EGB}}<\Omega_{\mathrm{GR}}$ and, consequently, a reduction in the spin precession frequency relative to the general relativity prediction. Furthermore, the characteristic $\rho^{-3}$ dependence shows that the EGB contribution is rapidly suppressed at large distances—where the general relativity result is smoothly recovered—while becoming increasingly significant as the particle approaches the compact object. The relative correction thus provides a convenient dimensionless quantity for identifying the radial region where higher-curvature effects may become significant in spin dynamics. The figure below shows this analysis
\begin{figure}[H]
\centering
\includegraphics[scale=0.52]{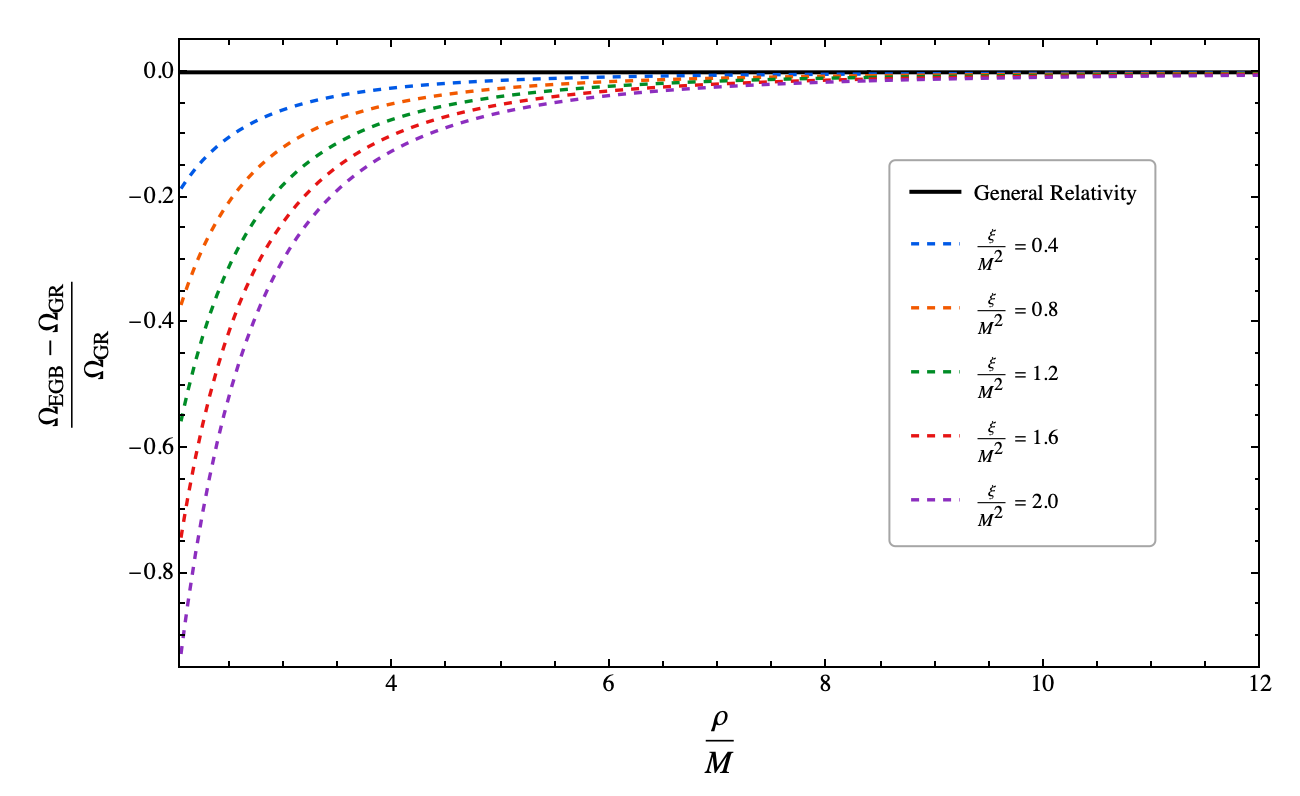}
\caption{\label{fig:spincorrection}Relative Einstein-Gauss-Bonnet correction to the spin-precession frequency as a function of the dimensionless radial coordinate $\rho/M$.}
\end{figure}
Figure \ref{fig:spincorrection} illustrates the radial dependence of the relative correction to the spin precession frequency for different values of the dimensionless coupling $\xi/M^{2}$. The horizontal black line represents the General Relativity (GR) limit, where the relative deviation vanishes identically, while the dashed curves correspond to predictions from EGB theory. All EGB curves lie below the GR reference for positive values of $\xi$, reflecting the reduction in precession frequency induced by the Gauss-Bonnet contribution. The deviation becomes progressively larger as $\rho/M$ decreases and $\xi/M^{2}$ increases, in accordance with the scaling law $-4(\xi/M^{2})(M/\rho)^{3}$. Conversely, at sufficiently large radial distances, all curves rapidly approach the GR limit, demonstrating that the higher-curvature correction becomes negligible in the weak-field region. This behavior is particularly relevant from an astrophysical perspective, as it indicates that spin precession observables are potentially more sensitive to EGB coupling in the vicinity of compact objects, whereas systems probing comparatively large orbital radii should exhibit strongly suppressed deviations from GR.

The connection between the quantum operator dynamics and its semiclassical interpretation can be established through Ehrenfest's theorem \cite{liboff2003introductory,merzbacher1998quantum,landau1958quantum}. For an operator $\mathcal{A}$ with no explicit time dependence, its expectation value satisfies
\begin{equation}
    \frac{d}{dt}\langle \mathcal{A}\rangle
    =\langle[H,\mathcal{A}]\rangle .
\end{equation}
Therefore, the Heisenberg equations derived above directly determine the evolution of the expectation values,
\begin{equation}
    \frac{d}{dt}\langle\boldsymbol{\rho}\rangle
    =\langle\mathbf{v}\rangle,
    \qquad
    \frac{d}{dt}\langle\boldsymbol{p}\rangle
    =\langle\mathbf{F}\rangle,
    \qquad
    \frac{d}{dt}\langle {\bf S}\rangle
    =\langle\dot{{\bf S}}\rangle .
\end{equation}
For sufficiently localized wave packets, these relations provide the semiclassical correspondence between the Foldy-Wouthuysen quantum dynamics and particle motion in the gravitational background. In particular, the EGB-dependent contributions to the velocity, force, and spin-precession operators generate the corresponding corrections to
their expectation-value dynamics.

\subsubsection{Orbital and total angular momentum}

The orbital angular momentum is not, in general, separately conserved in the presence of the spin-orbit interaction. From
\begin{equation}
\frac{d\mathbf{L}}{dt} = i[H_{\mathrm{FW}}, \mathbf{L}],
\label{D40}
\end{equation}
the rotationally invariant spin-independent terms do not contribute, while \(H_{\mathrm{SO}}\) gives
\begin{equation}
\frac{d\mathbf{L}}{dt} = 2C(\rho)\,\mathbf{S}\times\mathbf{L}.
\label{D41}
\end{equation}

Since \(\mathbf{S}\times\mathbf{L} = -\mathbf{L}\times\mathbf{S}\), Eqs.~(\ref{D28}) and (\ref{D41}) imply
\begin{equation}
\frac{d}{dt}(\mathbf{L} + \mathbf{S}) = 0.
\label{D42}
\end{equation}

\begin{figure}[H]
\centering
\includegraphics[scale=0.2]{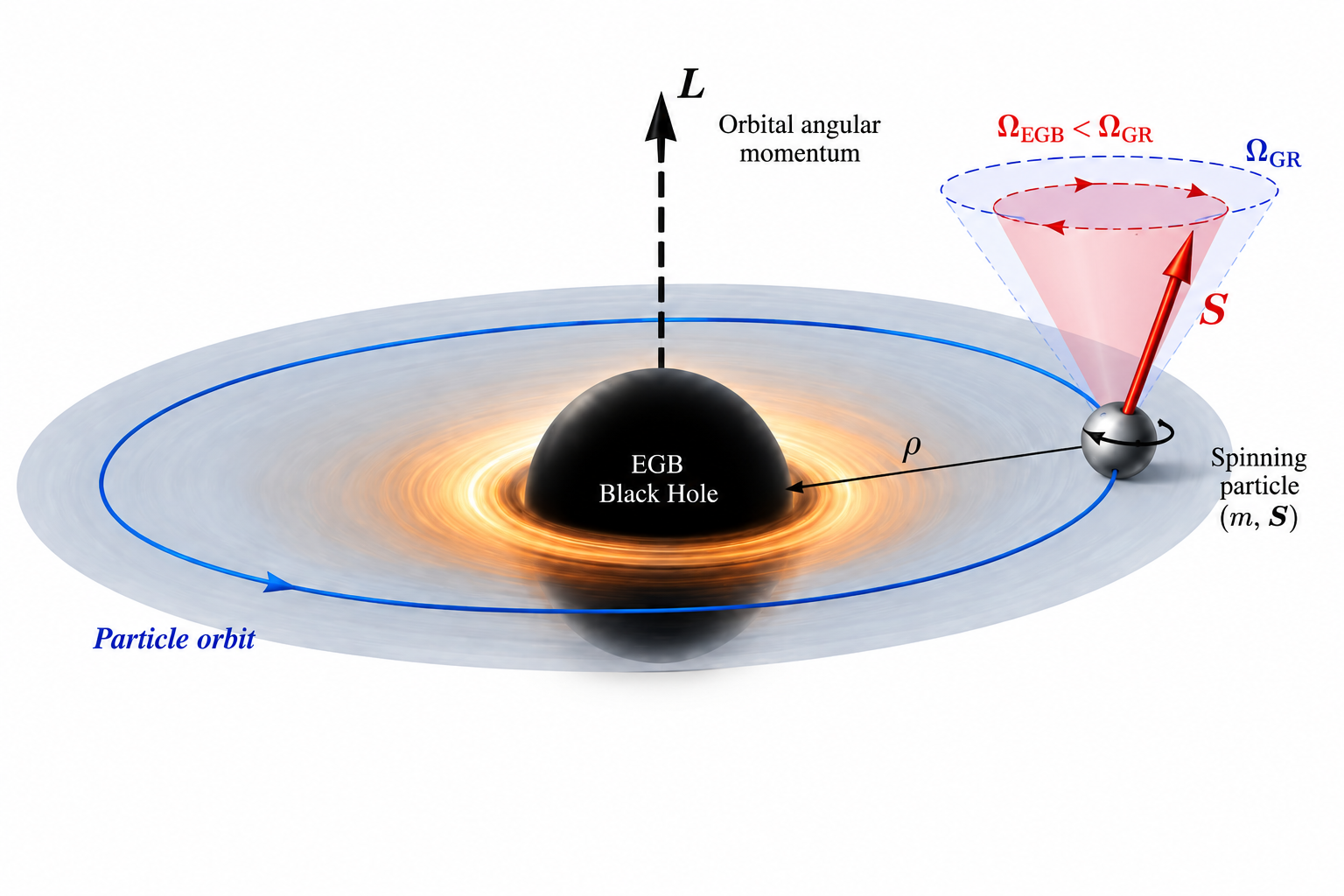}
\caption{\label{fig:spin} Schematic representation of the spin-precession dynamics of a spin-$1/2$ particle orbiting an Einstein-Gauss-Bonnet black hole.}
\end{figure}

Consequently, the total angular momentum
\begin{equation}
\mathbf{J} \equiv \mathbf{L} + \mathbf{S}
\label{D43}
\end{equation}
is conserved, as expected from the spherical symmetry of the gravitational background. The spin-orbit interaction therefore describes an exchange of angular momentum between the orbital and intrinsic degrees of freedom while preserving the total angular momentum. The figure (\ref{fig:spin})presents a schematic representation of the spin dynamics of a particle orbiting an Einstein–Gauss–Bonnet (EGB) black hole.

The orbital angular momentum $\mathbf{L}$ is perpendicular to the orbital plane, while the spin vector $\mathbf{S}$ precesses around this direction due to gravitational spin-orbit coupling. The precession cones illustrate the comparison between the General Relativity prediction, characterized by $\Omega_{\mathrm{GR}}$, and the dynamics modified by the Gauss–Bonnet contribution, $\Omega_{\mathrm{EGB}}$. For $\xi>0$, the EGB correction reduces the precession frequency relative to the GR result ($\Omega_{\mathrm{EGB}}<\Omega_{\mathrm{GR}}$), an effect that becomes increasingly significant as the particle approaches the compact object. The figure thus highlights the geometric interpretation of the EGB correction: the orbital trajectory provides the angular momentum that defines the precession axis, while the modification of the gravitational geometry alters the rate at which the spin evolves around that axis.
\subsubsection{Semiclassical limit}

As previously discussed in earlier sections about the Ehrenfest's theorem, the semiclassical orbital regime, where \(L \gg S\) (i.e., the orbital angular momentum is much larger than the spin), the backreaction of the spin dynamics on the orbital angular momentum may be neglected to leading order. In this limit, \(\mathbf{L}\) can be regarded as approximately fixed during the spin precession. Using
\begin{equation}
\mathbf{L} \simeq m\,\boldsymbol{\rho}\times\mathbf{v},
\label{D44}
\end{equation}
Eq.~(\ref{D29}) becomes
\begin{equation}
\boldsymbol{\Omega}_{\mathrm{EGB}} \simeq -\frac{3M}{2\rho^{3}}\left(1 - \frac{4\xi M}{\rho^{3}}\right)\boldsymbol{\rho}\times\mathbf{v}.
\label{D45}
\end{equation}

In the GR limit, \(\xi \rightarrow 0\), this expression reduces to
\begin{equation}
\boldsymbol{\Omega}_{\mathrm{GR}} \simeq -\frac{3M}{2\rho^{3}}\,\boldsymbol{\rho}\times\mathbf{v},
\label{D46}
\end{equation}
recovering the characteristic weak-field structure of the geodetic spin precession (the factor \(3/2\) is the well-known de Sitter precession coefficient for a particle in a Schwarzschild field).

\paragraph{Circular orbit approximation.} For an approximately circular orbit, one may further use
\begin{equation}
L \simeq m\rho^{2}\omega_{\mathrm{orb}},
\label{D47}
\end{equation}
which gives
\begin{equation}
\Omega_{\mathrm{EGB}} \simeq \frac{3M}{2\rho}\left(1 - \frac{4\xi M}{\rho^{3}}\right)\omega_{\mathrm{orb}}.
\label{D48}
\end{equation}
Here, $\omega_{\text{orb}}$ is the orbital angular frequency of the particle. An additional quantity of physical interest is the spin precession angle accumulated over a complete orbital revolution. While the precession frequency characterizes the instantaneous rate at which the spin vector rotates relative to the orbital motion, the accumulated angle measures the net spin rotation acquired after the particle completes an orbit. The accumulated spin-precession angle over one orbital period, \(T_{\mathrm{orb}} = 2\pi/\omega_{\mathrm{orb}}\), is 

\begin{equation}
\Delta\Phi_{\rm EGB}
\simeq
\frac{3\pi M}{\rho}
\left(
1-\frac{4\xi M}{\rho^{3}}
\right)
=
\Delta\Phi_{\rm GR}
-\frac{12\pi\xi M^{2}}{\rho^{4}},
\end{equation}
where $\Delta\Phi_{\rm GR}=3\pi M/\rho$ denotes the corresponding contribution from general relativity. The EGB modification thus produces an additional correction scaling as $\rho^{-4}$, which reduces the accumulated precession for positive $\xi$. This quantity is particularly useful because even a small modification in the instantaneous precession rate can, in principle, become more evident when the spin evolution is tracked over successive orbital revolutions. The figure below shows the behavior of the cumulative spin frequency.
\begin{figure}[H]
\centering
\includegraphics[scale=0.52]{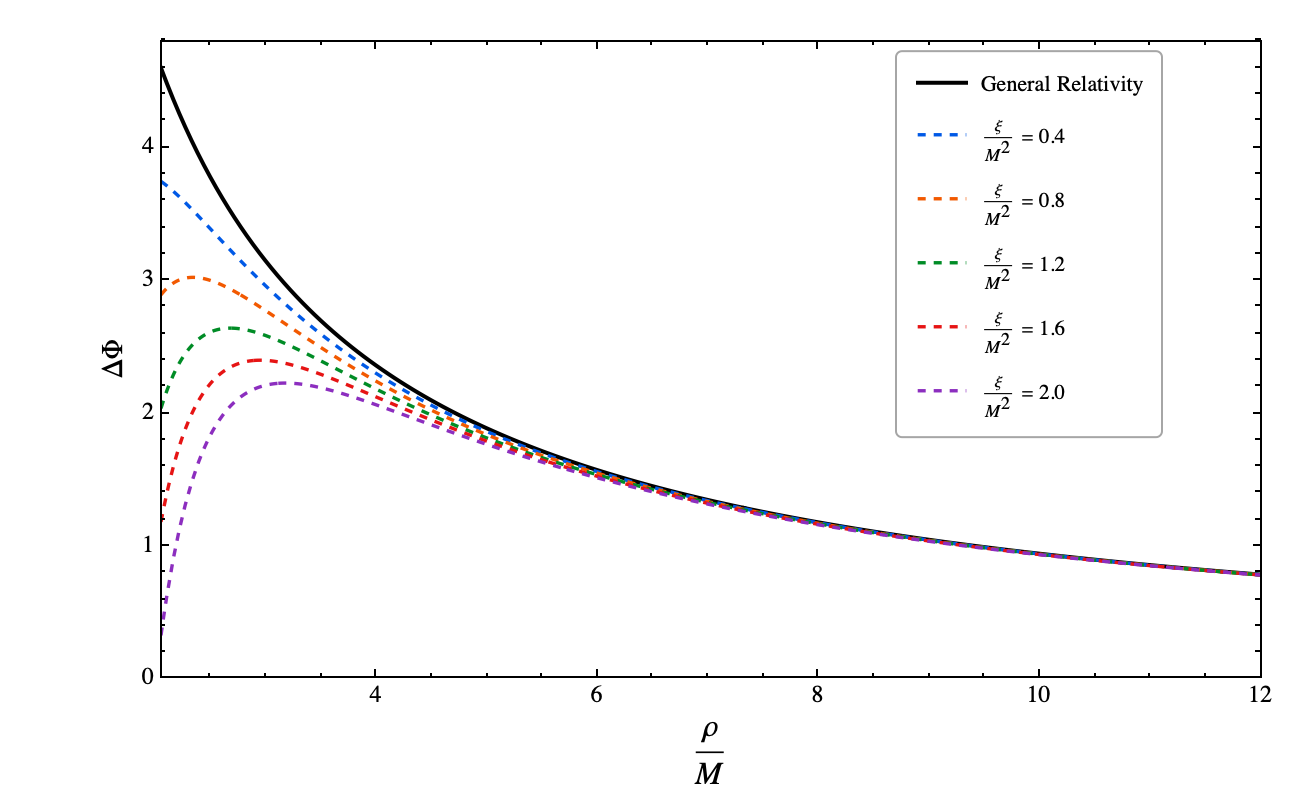}
\caption{\label{fig:acumuledspin}Accumulated spin-precession angle $\Delta\Phi$ as a function of the dimensionless radial coordinate $\rho/M$.}
\end{figure}
The figure~\ref{fig:acumuledspin} shows the spin precession angle accumulated per orbital revolution as a function of the dimensionless radial coordinate $\rho/M$. The solid black curve represents the general relativity prediction, while the dashed curves correspond to different values of the dimensionless Gauss-Bonnet coupling $\xi/M^{2}$. For positive values of $\xi$, the EGB curves remain below the GR result, demonstrating that the higher-curvature contribution reduces the total spin rotation accumulated during a complete orbit. The separation between the EGB and GR predictions becomes more pronounced at smaller radial distances, which is consistent with the characteristic $\rho^{-4}$ dependence of the Gauss-Bonnet contribution to the accumulated angle. Conversely, as $\rho/M$ increases, all curves rapidly converge to the GR prediction, reflecting the suppression of higher-curvature effects in the weak-field regime. This behavior demonstrates that the accumulated precession angle provides a direct and physically transparent observable quantity for assessing the influence of EGB coupling on spin transport around compact objects.
\subsection{Astrophysical sensitivity to the EGB coupling: the A0620-00 system}

As an astrophysically motivated application of the preceding results, we consider the stellar-mass black-hole binary A0620--00. This source is particularly suitable for the present analysis because its physical parameters have been accurately constrained from observations. Cantrell et al.~\cite{cantrell2010inclination} obtained an orbital inclination $i=51.0^{\circ}\pm0.9^{\circ}$, a black-hole mass $M_{\rm BH}=6.6\pm0.25\,M_{\odot}$, and a distance $D=1.06\pm0.12\,{\rm kpc}$. Moreover, the continuum-fitting analysis of Gou et al.~\cite{gou2010spin} yielded a comparatively small dimensionless spin parameter, $a_{*}=0.12\pm0.19$, with $a_{*}<0.49$ at the $3\sigma$ confidence level. The relatively small inferred spin makes A0620-00 a useful astrophysical benchmark for the approximately static and spherically symmetric gravitational background considered in the present work.

Rather than interpreting the following analysis as a direct observational bound, we use the measured mass of A0620-00 to quantify the prospective sensitivity of spin-precession measurements to the EGB coupling. From the relative precession correction, one obtains
\begin{equation}
|\xi|=
\frac{M^{2}}{4}
\left(\frac{\rho}{M}\right)^{3}
\left|
\frac{\Omega_{\rm EGB}-\Omega_{\rm GR}}
{\Omega_{\rm GR}}
\right|.
\end{equation}

For the A0620-00 system, taking $M_{\rm BH}=6.61\,M_{\odot}$ gives
\begin{equation}
M=\frac{GM_{\rm BH}}{c^{2}}
\simeq 9.76\times10^{3}\ {\rm m},
\end{equation}
and consequently
\begin{equation}
M^{2}\simeq 9.53\times10^{7}\ {\rm m}^{2}.
\end{equation}
Using
\begin{equation}
|\xi|=
\frac{M^{2}}{4}
\left(\frac{\rho}{M}\right)^{3}
\left|
\frac{\Omega_{\rm EGB}-\Omega_{\rm GR}}
{\Omega_{\rm GR}}
\right|,
\end{equation}
one can estimate the values of the Gauss--Bonnet coupling that could be probed for different orbital radii and relative sensitivities in the spin-precession frequency. For instance, at $\rho=10M$, one obtains
\begin{equation}
|\xi|
\simeq
2.38\times10^{10}
\left|
\frac{\Omega_{\rm EGB}-\Omega_{\rm GR}}
{\Omega_{\rm GR}}
\right|
{\rm m}^{2}.
\end{equation}

Thus, relative sensitivities of $10^{-2}$, $10^{-3}$, and $10^{-4}$ they must correspond to different values of $\xi$. It is important to emphasize that the orbital radial considered below, $\rho=10M$--$30M$, do not correspond to the orbital separation of the stellar companion in the A0620--00 binary. The observed system is used here to fix a realistic stellar-mass black-hole scale, whereas $\rho$ denotes the orbital radius of the spin-$1/2$ test particle described by our effective Hamiltonian. The resulting estimates should therefore be interpreted as source-calibrated prospective sensitivities rather than as constraints derived from the observed binary orbit. These values can be seen in the table below.

\begin{table}[H]
\centering
\caption{Prospective sensitivity to the EGB coupling $\xi$ for the
A0620--00 system, assuming $M_{\rm BH}=6.61\,M_{\odot}$.}
\label{tab:xi_A0620}
\begin{tabular}{c|ccc}
\hline\hline
$\rho/M$
&
$|\delta_{\Omega}|=10^{-2}$
&
$|\delta_{\Omega}|=10^{-3}$
&
$|\delta_{\Omega}|=10^{-4}$
\\
\hline
10
&
$2.38\times10^{8}$
&
$2.38\times10^{7}$
&
$2.38\times10^{6}$
\\
20
&
$1.91\times10^{9}$
&
$1.91\times10^{8}$
&
$1.91\times10^{7}$
\\
30
&
$6.43\times10^{9}$
&
$6.43\times10^{8}$
&
$6.43\times10^{7}$
\\
\hline\hline
\end{tabular}
\\[1mm]
{\small All values of $\xi$ are expressed in ${\rm m}^{2}$.}
\end{table}

The discrete estimates presented in Table~\ref{tab:xi_A0620} can be extended continuously over the orbital region considered here. It is instructive to compare these prospective sensitivities with existing phenomenological constraints on four-dimensional Einstein-Gauss-Bonnet gravity. Clifton et al.~\cite{clifton2020observational} investigated a regularized formulation of the theory using weak-field, cosmological, and black-hole observations. Their analysis indicates an overall positive-coupling scale approximately bounded by $0\lesssim\xi\lesssim10^{8}\,{\rm m}^{2}$, while weak-field tests alone lead to less restrictive constraints. This comparison is particularly interesting in the present context, since our estimates for A0620--00 show that a relative spin-precession sensitivity at the level of $10^{-3}$--$10^{-4}$ could probe EGB couplings in the approximate range $10^{6}$--$10^{8}\,{\rm m}^{2}$ for sufficiently compact test-particle orbits. The comparison should nevertheless be regarded as indicative rather than as a direct translation of existing bounds, since constraints on the Gauss--Bonnet coupling depend on the particular four-dimensional realization of the theory.

Figure below shows the prospective sensitivity to the Gauss--Bonnet coupling $\xi$ for the stellar-mass black-hole system A0620-00 as a function of the dimensionless orbital radius $\rho/M$. The three curves correspond to assumed relative sensitivities in the spin-precession frequency of $|\delta_{\Omega}|=10^{-2}$, $10^{-3}$, and $10^{-4}$, where $\delta_{\Omega}\equiv (\Omega_{\rm EGB}-\Omega_{\rm GR})/\Omega_{\rm GR}$. For a fixed observational sensitivity, the value of $\xi$ that can be probed increases rapidly with the orbital radius, following the scaling $\xi\propto(\rho/M)^{3}$. Consequently, trajectories closer to the compact object provide a substantially enhanced sensitivity to the Gauss-Bonnet coupling.

\begin{figure}[H]
\centering
\includegraphics[scale=0.52]{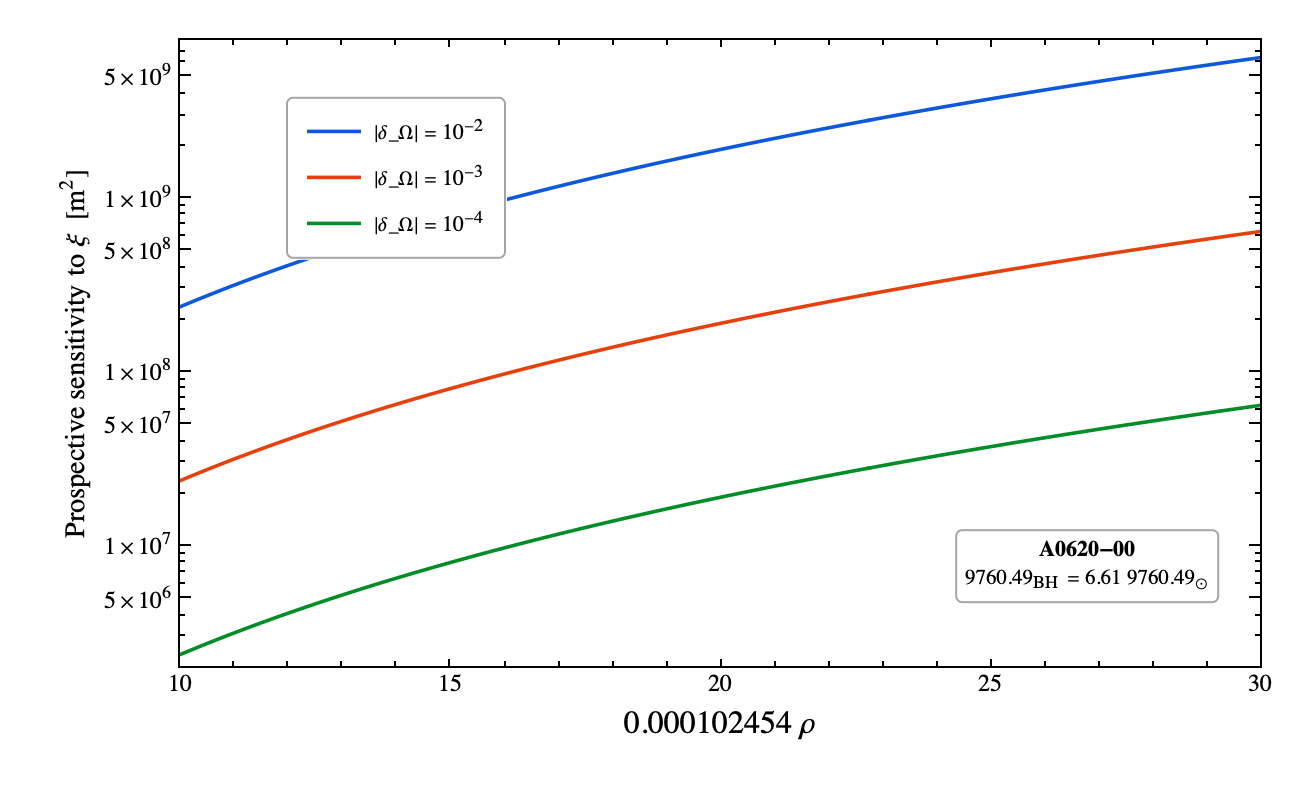}
\caption{\label{fig:sensity} Prospective sensitivity to the Einstein-Gauss-Bonnet coupling parameter $\xi$ for the compact binary system A0620-00, shown as a function of the radial coordinate $\rho$.}
\end{figure}

The figure (\ref{fig:sensity}) therefore illustrates two complementary aspects of the phenomenology. First, the strong $(M/\rho)^{3}$ dependence makes the near-field region of stellar-mass compact objects particularly favorable for probing higher-curvature corrections. Second, improved precision in spin-precession measurements directly extends the sensitivity toward smaller values of $\xi$. The curves should, however, be interpreted as prospective sensitivity estimates rather than present observational bounds, since the intrinsic spin precession considered in the present analysis has not yet been measured for a test spin-$1/2$ particle orbiting A0620--00.

The analysis presented above establishes a direct connection between the nonrelativistic quantum dynamics of spin-$1/2$ particles and the phenomenology of compact objects in Einstein-Gauss-Bonnet gravity. In particular, the Foldy-Wouthuysen and Heisenberg formulations allow the higher-curvature contribution to be identified directly at the level of the spin dynamics, leading to a characteristic modification of the precession frequency that scales as $\delta_{\Omega}\propto\xi M/\rho^{3}$. This strong radial dependence makes stellar-mass compact objects especially interesting laboratories for probing such corrections. Using the measured mass of A0620--00, we have shown that relative spin-precession sensitivities at the level of $10^{-3}$--$10^{-4}$ would probe values of the Gauss--Bonnet coupling extending approximately into the $10^{6}$--$10^{8}\,{\rm m}^{2}$ range, depending on the orbital radius. Although these values should presently be interpreted as prospective sensitivities rather than direct observational bounds, they demonstrate that the quantum spin dynamics derived here can be translated into a quantitatively testable astrophysical framework. More generally, our results provide
a bridge between the effective quantum dynamics of fermions in modified gravitational backgrounds and precision tests of gravity around compact objects, suggesting spin-precession observables as an additional channel through which higher-curvature effects may be investigated alongside conventional orbital and gravitational wave probes.

\section{Final Remarks} 
\label{sec:conclusions}
In this work, I have investigated the non relativistic quantum dynamics of a spin-$1/2$ particle in a static and spherically symmetric Einstein--Gauss--Bonnet (EGB) gravitational background. Starting from the curved-spacetime Dirac theory, we constructed the corresponding Hermitian Hamiltonian in isotropic coordinates and performed a Foldy--Wouthuysen transformation to identify the low-energy gravitational dynamics. The resulting Hamiltonian systematically separates the conventional non relativistic and general-relativistic contributions from the higher-curvature corrections controlled by the Gauss--Bonnet coupling $\xi$. In addition to modifications of the gravitational and kinetic sectors, the effective theory contains spin--orbit and Darwin-type terms, showing explicitly how the EGB geometry enters the quantum dynamics of fermionic degrees of freedom. This construction extends the well-established FW description of fermions in gravitational fields \cite{obukhov2001spin,obukhov2013spin,jentschura2013nonrelativistic} to the higher-curvature background considered here.

The Heisenberg equations provide a direct dynamical interpretation of the effective Hamiltonian. We derived the velocity and force operators and showed that the EGB contribution affects both the translational and spin-dependent sectors of the motion. In particular, the effective force contains modified radial and momentum-dependent gravitational
terms together with spin-dependent contributions associated with the spin-orbit interaction. Through the Ehrenfest theorem, the expectation values of these operators establish the connection between the quantum description and the corresponding semiclassical dynamics. The resulting picture therefore goes beyond a purely geodesic analysis by retaining
the intrinsic spin degree of freedom and its coupling to the modified gravitational background.

The spin dynamics provides one of the main physical results of the present analysis. The EGB correction modifies the gravitational spin-orbit interaction and consequently changes the precession of the fermionic spin. For positive $\xi$, the higher-curvature contribution therefore reduces the spin-precession frequency with respect to its GR value,
while the characteristic $(M/\rho)^{3}$ dependence ensures the continuous recovery of GR at large distances. The accumulated precession angle provides a complementary manifestation of the same effect, with the Gauss--Bonnet contribution scaling as $\delta\Phi_{\rm GB}\propto-\xi M^{2}/\rho^{4}$. Consequently, even when the instantaneous modification is small, repeated orbital cycles may enhance the accumulated difference between the EGB and GR spin evolution, provided that the orbital configuration remains sufficiently stable.

To connect these quantum-gravitational results with astrophysical scales, we considered the stellar-mass black-hole system A0620-00. Its accurately determined mass and comparatively small inferred spin make it a particularly useful benchmark for the approximately static and spherically symmetric geometry adopted in our analysis
\cite{cantrell2010inclination,gou2010spin}. Using $M_{\rm BH}\simeq6.61\,M_{\odot}$, corresponding in geometrized units to $M\simeq9.76\,{\rm km}$ and
$M^{2}\simeq9.53\times10^{7}\,{\rm m}^{2}$, we translated the spin-precession correction into a prospective sensitivity to the Gauss-Bonnet coupling. Relative sensitivities of
$10^{-3}$--$10^{-4}$ lead to characteristic values of $\xi$ in the approximate range $10^{6}$--$10^{8}\,{\rm m}^{2}$ for the compact orbital configurations considered here. In particular, at $\rho=20M$, a relative sensitivity of $10^{-4}$ corresponds to $|\xi|\simeq1.91\times10^{7}\,{\rm m}^{2}$.

These scales are particularly interesting when compared with existing phenomenological studies of four-dimensional EGB gravity, which have used weak-field tests, binary systems, gravitational observations, and compact objects to constrain Gauss--Bonnet modifications
\cite{clifton2020observational,charmousis2022astrophysical}. Nevertheless, the estimates obtained here should not be interpreted as new observational bounds on $\xi$. Rather, they quantify the prospective sensitivity of the fermionic spin-precession channel for a real astrophysical mass scale. Moreover, direct numerical comparisons between constraints obtained in different four-dimensional Gauss-Bonnet constructions must be made with care, since the precise meaning and normalization of the coupling can depend on the underlying
regularization or scalar-tensor realization \cite{fernandes2020derivation,hennigar2020lower,lu2020horndeski,aoki2020consistent,fernandes2022Review}.

The present results therefore suggest a complementary perspective on tests of higher-curvature gravity. Whereas most compact-object probes are based on orbital dynamics, wave propagation, accretion phenomena, or gravitational radiation, the formalism developed here identifies the quantum spin degree of freedom itself as a carrier of information about the gravitational background. The combination of the Foldy-Wouthuysen representation, Heisenberg dynamics, and the Ehrenfest correspondence provides a systematic route from the
relativistic Dirac equation to potentially testable spin-dependent signatures of modified gravity. In this sense, fermionic spin precession constitutes an additional channel through which higher-curvature effects may be investigated alongside conventional strong-gravity observables.

Several extensions follow naturally from this framework. The inclusion of rotation would allow the analysis to be generalized to stationary axisymmetric EGB compact objects and would make it possible to study the interplay between frame dragging and the higher-curvature spin-orbit interaction. Strong-field treatments beyond the present
perturbative FW expansion, as well as noncircular and time-dependent orbital configurations, would further enlarge the range of applicability. Ultimately, combining spin-sensitive quantum dynamics with increasingly precise compact-object observations may provide a new route for testing the gravitational interaction at the interface between quantum theory and strong-field gravity.
\section*{Acknowledgements} 
The author acknowledges the Graduate Program in Physics at the Federal University of Campina Grande.
\appendix{}
\appendix
\section{The Einstein-Gauss-Bonnet metric in isotropic coordinates}
\label{app:EGB_isotropic}

In this appendix we provide a detailed derivation of the Einstein-Gauss-Bonnet (EGB) metric in isotropic coordinates, up to first order in the Gauss-Bonnet coupling $\xi$ (which corresponds to order $M^2$ in the mass expansion). This form is essential for the Hamiltonian formulation of the Dirac equation, since isotropic coordinates render the spatial part of the metric manifestly conformally flat.

\subsection{The starting point: Schwarzschild-Gauss-Bonnet coordinates}

We begin with the standard spherically symmetric solution of the EGB field equations in $D$-dimensional spacetime, written in Schwarzschild--Gauss-Bonnet coordinates~\cite{Boulware1985}:
\begin{equation}
ds^2 = -f(r)\,dt^2 + \frac{dr^2}{f(r)} + r^2\,d\Omega_{D-2}^2 , 
\label{eq:EGB_schw}
\end{equation}
where $d\Omega_{D-2}^2$ is the metric on the unit $(D-2)$-sphere and
\begin{equation}
f(r) = 1 + \frac{r^2}{2\xi} \left[ 1 \pm \sqrt{1 + \frac{8\xi M}{r^3}} \right].
\label{eq:f_EGB_exact}
\end{equation}
Here $M$ is the mass parameter (related to the ADM mass) and $\xi$ is the Gauss-Bonnet coupling constant with dimensions of $[L]^2$. The minus sign in front of the square root is selected because it recovers the usual Schwarzschild solution in the limit $\xi \to 0$. Since we are interested in the weak-field regime (small $M$) and in the leading-order correction due to the EGB term, we expand the square root in Eq.~\eqref{eq:f_EGB_exact} for $\xi M / r^3 \ll 1$:
\begin{equation}
\sqrt{1 + \frac{8\xi M}{r^3}} 
= 1 + \frac{4\xi M}{r^3} - \frac{8\xi^2 M^2}{r^6} + \mathcal{O}(\xi^3 M^3).
\label{eq:sqrt_exp}
\end{equation}
Substituting (\ref{eq:sqrt_exp}) into (\ref{eq:f_EGB_exact}) and keeping terms up to order $M^2$ (and hence linear in $\xi$, since $\xi$ already multiplies $M^2$), we obtain

\begin{align}
f(r) &= 1 + \frac{r^2}{2\xi} \left[ 1 - \left( 1 + \frac{4\xi M}{r^3} - \frac{8\xi^2 M^2}{r^6} \right) \right] \nonumber \\
&= 1 - \frac{2M}{r} + \frac{4\xi M^2}{r^4} + \mathcal{O}(M^3, \xi^2 M^3).
\label{eq:f_expanded}
\end{align}

Notice that the term $4\xi M^2/r^4$ is the unique modification introduced by the Gauss-Bonnet sector at this perturbative order. We now seek a coordinate change $r = r(\rho)$ such that the spatial part of the metric becomes proportional to the flat Euclidean metric in $D-1$ dimensions, i.e.
\begin{equation}
ds^2 = -F(\rho)\,dt^2 + H(\rho)\bigl(d\rho^2 + \rho^2\,d\Omega_{D-2}^2\bigr),
\label{eq:isotropic_ansatz}
\end{equation}
where $F(\rho)$ and $H(\rho)$ are functions to be determined. The condition for the absence of cross terms $dt\,d\rho$ is automatically satisfied by spherical symmetry, and the relation between the two coordinate systems is obtained by comparing the radial parts of (\ref{eq:EGB_schw}) and (\ref{eq:isotropic_ansatz}):
\begin{equation}
\frac{H(\rho)}{F(\rho)} \left(\frac{d\rho}{dr}\right)^2 = \frac{1}{f(r)}.
\label{eq:radial_condition}
\end{equation}

Instead of solving this nonlinear ODE exactly, we adopt a perturbative ansatz for the transformation that generalises the well-known isotropic form of Schwarzschild:

\begin{equation}
r(\rho) = \rho \left(1 + \frac{M}{2\rho}\right)^2 - \frac{4\xi M^2}{3\rho^3} + \mathcal{O}(M^3, \xi^2 M^3).
\label{eq:r_rho_ansatz}
\end{equation}
The first term is the standard Schwarzschild–isotropic relation, while the second term (proportional to $\xi M^2$) encodes the leading EGB deformation. The numerical coefficient $-4/3$ is fixed by requiring that the metric (\ref{eq:isotropic_ansatz}) satisfies the field equations (or, equivalently, by demanding that the cross term vanishes identically after substitution). For completeness, we invert (\ref{eq:r_rho_ansatz}) to express $1/r$ in terms of $\rho$:
\begin{align}
\frac{1}{r} &= \frac{1}{\rho} \left(1 + \frac{M}{2\rho}\right)^{-2} 
\left[ 1 - \frac{4\xi M^2}{3\rho^4} \right]^{-1} \nonumber \\
&= \frac{1}{\rho} - \frac{M}{\rho^2} + \frac{M^2}{4\rho^3} + \frac{4\xi M^2}{3\rho^5} + \mathcal{O}(M^3),
\label{eq:1_r_exp}
\end{align}
and similarly for higher inverse powers. In particular, we shall need
\begin{equation}
\frac{1}{r^4} = \frac{1}{\rho^4} - \frac{2M}{\rho^5} + \mathcal{O}(M^2).
\label{eq:1_r4_exp}
\end{equation}

Using (\ref{eq:r_rho_ansatz})--(\ref{eq:1_r4_exp}), we can now compute $F(\rho)$ and $H(\rho)$ explicitly. From $F(\rho) = f(r(\rho))$ and the expansion (\ref{eq:f_expanded}), we get
\begin{align}
F(\rho) &= 1 - \frac{2M}{r} + \frac{4\xi M^2}{r^4} \nonumber \\
&= 1 - 2M\left( \frac{1}{\rho} - \frac{M}{\rho^2} + \frac{M^2}{4\rho^3} \right) 
+ \frac{4\xi M^2}{\rho^4} + \mathcal{O}(M^3) \nonumber \\
&= 1 - \frac{2M}{\rho} + \frac{2M^2}{\rho^2} + \frac{4\xi M^2}{\rho^4} + \mathcal{O}(M^3, \xi^2 M^3).
\label{eq:F_iso}
\end{align}

The spatial factor is obtained from the radial part of (\ref{eq:isotropic_ansatz}). Comparing with (\ref{eq:EGB_schw}) gives
\begin{equation}
H(\rho) = \left( \frac{r}{\rho} \right)^2 \left( \frac{dr}{d\rho} \right)^{-2} f(r).
\label{eq:H_def}
\end{equation}
A more direct and equivalent route is to use the relation $H(\rho) = (r/\rho)^2$. From (\ref{eq:r_rho_ansatz}), we compute
\begin{align}
\frac{r}{\rho} &= \left(1 + \frac{M}{2\rho}\right)^2 - \frac{4\xi M^2}{3\rho^4} \nonumber \\
&= 1 + \frac{M}{\rho} + \frac{M^2}{4\rho^2} - \frac{4\xi M^2}{3\rho^4} + \mathcal{O}(M^3).
\label{eq:r_over_rho}
\end{align}
Squaring this expression and keeping only terms up to order $M^2$ (and linear in $\xi$), we find
\begin{align}
H(\rho) &= \left( \frac{r}{\rho} \right)^2 \nonumber \\
&= 1 + \frac{2M}{\rho} - \frac{8\xi M^2}{3\rho^4} + \mathcal{O}(M^3, \xi^2 M^3).
\label{eq:H_iso}
\end{align}

\subsection{Final form of the metric}

Collecting the results (\ref{eq:F_iso}) and (\ref{eq:H_iso}), the EGB metric in isotropic coordinates reads, up to first order in the EGB correction,
\begin{equation}
\begin{aligned}
ds^2 = &-\left( 1 - \frac{2M}{\rho} + \frac{4\xi M^2}{\rho^4} \right) dt^2 \\
&+ \left( 1 + \frac{2M}{\rho} - \frac{8\xi M^2}{3\rho^4} \right)
\left( d\rho^2 + \rho^2\,d\Omega_{D-2}^2 \right) \\
&+ \mathcal{O}(M^3, \xi^2 M^3).
\end{aligned}
\label{eq:EGB_isotropic_final}
\end{equation}

Several remarks are in order:

\begin{enumerate}
\item \textbf{Limiting cases.} When $\xi \to 0$, (\ref{eq:EGB_isotropic_final}) reduces to the standard Schwarzschild metric in isotropic coordinates expanded to order $M$, i.e.
$$
F(\rho) = 1 - \frac{2M}{\rho} , \qquad
H(\rho) = 1 + \frac{2M}{\rho} ,
$$
which is precisely the expansion of the exact relation $F = [(1 - M/2\rho)/(1 + M/2\rho)]^2$ and $H = (1 + M/2\rho)^4$.

\item \textbf{Sign of the EGB correction.} The Gauss-Bonnet term introduces opposite signs in the temporal and spatial sectors: $+4\xi M^2/\rho^4$ in $F(\rho)$ and $-8\xi M^2/(3\rho^4)$ in $H(\rho)$. This relative sign is responsible for the characteristic screening effects at short distances and modifies the effective potential in the Dirac Hamiltonian.

\item \textbf{Coordinate relabelling.} In the main text, after performing this transformation, we drop the hat and simply write $\rho \to r$ to avoid heavy notation. The reader should bear in mind that, from this point onward, $r$ denotes the isotropic radial coordinate.

\item \textbf{Validity.} The expansion is valid for $\xi M / r^3 \ll 1$ and $M/r \ll 1$. This covers the weak-field and small-coupling regime, which is precisely the domain where the perturbative Hamiltonian treatment is justified.
\end{enumerate}

The importance of the isotropic form (\ref{eq:EGB_isotropic_final}) lies in the fact that the spatial metric is now conformally flat:
\begin{equation}
g_{ij} = H(\rho)\,\delta_{ij}, \qquad i,j=1,\dots,D-1.
\end{equation}
Consequently, the Dirac operator can be written in terms of the ordinary flat-space gamma matrices and the conformal factor $H$. In particular, the square root of the determinant simplifies to $\sqrt{-g} = \sqrt{F}\, H^{(D-1)/2}$, and the spin connection acquires simple derivatives of $\ln H$. This structure is exploited in Sec.~\ref{app:FW_general} to derive the modified Dirac Hamiltonian in EGB gravity. This concludes the derivation of the EGB metric in isotropic coordinates.

\section{Foldy-Wouthuysen transformation for a general static spherically symmetric metric}
\label{app:FW_general}

In this appendix, we present a self-contained derivation of the non-relativistic Foldy-Wouthuysen (FW) Hamiltonian for a Dirac particle propagating in an arbitrary static and spherically symmetric spacetime. The metric is written in isotropic coordinates as
\begin{equation}
ds^2 = -V^2\,dt^2 + W^2\,d\mathbf{x}^2,
\label{C1}
\end{equation}
where \(V\) and \(W\) are arbitrary positive functions of the radial coordinate. The derivation is carried out to order \(1/m^2\) in the non-relativistic expansion, which is sufficient to capture all gravitational corrections up to the spin-orbit and Darwin sectors. The results obtained here are then particularised in the main text to the Einstein--Gauss--Bonnet (EGB) and Schwarzschild backgrounds.

\subsection{The Hermitian Dirac Hamiltonian}

The Dirac equation in a curved spacetime can be cast into a Schr\"odinger-like form through a suitable field redefinition that absorbs the spin connection and renders the scalar product flat \cite{obukhov2001spin}. After this procedure, the Hermitian Dirac Hamiltonian takes the compact form

\begin{equation}
H_{\rm D} = \beta m V + \frac{1}{2}\left\{\mathcal{F},\, \boldsymbol{\alpha}\cdot\mathbf{p}\right\},
\label{C2}
\end{equation}
where \(\mathbf{p} = -i\boldsymbol{\nabla}\) is the momentum operator, \(\{\,,\}\) denotes the anticommutator, and we have defined the dimensionless function
\begin{equation}
\mathcal{F} \equiv \frac{V}{W}.
\label{C3}
\end{equation}

The Dirac matrices satisfy the standard Clifford algebra \(\{\alpha^i,\alpha^j\} = 2\delta^{ij}\) and \(\{\alpha^i,\beta\}=0\), with \(\beta^2 = \mathbb{I}\).
To apply the FW transformation, we decompose the Hamiltonian into even and odd parts with respect to \(\beta\):
\begin{equation}
H_{\rm D} = \beta m + \mathcal{E} + \mathcal{O},
\label{C4}
\end{equation}
where the even \((\mathcal{E})\) and odd \((\mathcal{O})\) operators satisfy
\begin{equation}
[\beta,\mathcal{E}] = 0, \qquad \{\beta,\mathcal{O}\} = 0.
\label{C5}
\end{equation}
For the metric (\ref{C1}), these operators are explicitly given by
\begin{align}
\mathcal{E} &= \beta m (V - 1), \label{C6} \\
\mathcal{O} &= \frac{1}{2}\left\{F,\, \boldsymbol{\alpha}\cdot\mathbf{p}\right\}. \label{C7}
\end{align}
We note that \(\mathcal{E}\) contains a term proportional to \(\beta\) and is therefore even, while \(\mathcal{O}\) contains an odd number of \(\alpha\)-matrices and is odd. The FW transformation is implemented through the unitary operator
\begin{equation}
U = e^{iS},
\label{C8}
\end{equation}
where the generator \(S\) is chosen order by order in the \(1/m\) expansion to systematically remove the odd operators. Following the standard procedure~\cite{foldy1950dirac, bjorken1965relativistic}, we write
\begin{equation}
S = S_1 + S_2 + S_3 + \cdots,
\label{C9}
\end{equation}
with
\begin{align}
S_1 &= -\frac{i}{2m}\beta\mathcal{O}, \label{C10} \\
S_2 &= -\frac{i}{4m^2}\beta[\mathcal{O},\mathcal{E}], \label{C11} \\
S_3 &= \frac{i}{3m^3}\beta\mathcal{O}^3. \label{C12}
\end{align}
The transformed Hamiltonian is obtained via the Baker-Campbell-Hausdorff expansion:
\begin{equation}
H' = e^{iS}H_{\rm D}e^{-iS} = H_{\rm D} + i[S,H_{\rm D}] + \frac{i^2}{2!}[S,[S,H_{\rm D}]] + \cdots .
\label{C13}
\end{equation}
After a lengthy but straightforward calculation, keeping terms up to order \(1/m^2\) in the even sector, the positive-energy FW Hamiltonian is found to be
\begin{equation}
\mathcal{H}_{FW}=\beta m+\mathcal{E}+\frac{\beta}{2m}\mathcal{O}^{2}-\frac{1}{8m^{2}}[\mathcal{O},[\mathcal{O},\mathcal{E}]]+\cdots .
\label{C14}
\end{equation}
The first line contains the free-particle Hamiltonian and the scalar potential; the second line gives the \(1/m\) corrections (kinetic, spin-orbit, and Darwin); and the third line contains the \(1/m^2\) corrections (including the gravitational modification of the \(\mathbf{p}^4\) term and higher-order spin interactions).

We now evaluate each term in Eq.~(\ref{C14}) in terms of the functions \(V(\rho)\), \(W(\rho)\), and their derivatives. To organise the calculation, we introduce the following auxiliary functions:
\begin{align}
v &\equiv V - 1, \label{C15} \\
f &\equiv F - 1 = \frac{V}{W} - 1, \label{C16} \\
k &\equiv \frac{V}{W^2} - 1, \label{C17} \\
\chi &\equiv 2f - v. \label{C18}
\end{align}
These combinations naturally appear in the various sectors of the Hamiltonian. The first correction is simply
\begin{equation}
\mathcal{E} = \beta m v.
\label{C19}
\end{equation}
This gives the classical gravitational potential energy. The square of the odd operator yields
\begin{equation}
\mathcal{O}^2 = \frac{1}{4}\left\{F,\,\boldsymbol{\alpha}\cdot\mathbf{p}\right\}^2.
\label{C20}
\end{equation}
Using the identity \((\boldsymbol{\alpha}\cdot\mathbf{p})^2 = \mathbf{p}^2\) and the fact that \(F\) is a function of \(\rho\), we obtain
\begin{equation}
\mathcal{O}^2 = F^2\,\frac{\mathbf{p}^2}{2} - \frac{i}{4}F\,\boldsymbol{\alpha}\cdot\boldsymbol{\nabla}F\cdot\boldsymbol{\alpha}\,\mathbf{p} + \mathcal{O}(\hbar^2).
\label{C21}
\end{equation}
More precisely, after symmetrisation to ensure Hermiticity,
\begin{equation}
\mathcal{O}^2 = \frac{1}{2}F^2\,\mathbf{p}^2 + \frac{1}{4}\boldsymbol{\alpha}\cdot\left\{\boldsymbol{\nabla}F,\,\boldsymbol{\alpha}\cdot\mathbf{p}\right\}.
\label{C22}
\end{equation}
The second term contains a spin-dependent part after using the identity \(\alpha^i\alpha^j = \delta^{ij} + i\epsilon^{ijk}\Sigma^k\) with \(\Sigma^k = \text{diag}(\sigma^k,\sigma^k)\). This will be evaluated below.

Using Eqs.~(\ref{C6}) and (\ref{C7}),
\begin{equation}
[\mathcal{O},\mathcal{E}] = \beta m\left[\frac{1}{2}\left\{F,\,\boldsymbol{\alpha}\cdot\mathbf{p}\right\},\,V-1\right].
\label{C23}
\end{equation}
Since \(V\) and \(F\) are radial functions, this simplifies to
\begin{equation}
[\mathcal{O},\mathcal{E}] = -\frac{i\beta m}{2}\left( \boldsymbol{\alpha}\cdot\boldsymbol{\nabla}F\,V + F\,\boldsymbol{\alpha}\cdot\boldsymbol{\nabla}V \right).
\label{C24}
\end{equation}
This term contributes to the spin-orbit interaction after the next commutator is evaluated. The \(1/m\) gravitational corrections are contained in
\begin{equation}
\frac{1}{4m}\left\{\mathcal{O}^2 - [\mathcal{O},\mathcal{E}]\right\}.
\label{C25}
\end{equation}
After substituting Eqs.~(\ref{C22}) and (\ref{C24}) and performing the spin algebra, we obtain
\begin{equation}
\frac{1}{4m}\left\{\mathcal{O}^2 - [\mathcal{O},\mathcal{E}]\right\}
= \frac{1}{8m}\left\{4F^2 - 2m(V-1),\,\mathbf{p}^2\right\} + H_{\rm SO} + H_{\rm Darwin},
\label{C26}
\end{equation}
where the spin-orbit and Darwin terms will be isolated below. Collecting all contributions and expressing the results in terms of the functions defined in Eqs.~(\ref{C15})--(\ref{C18}), the positive-energy FW Hamiltonian can be written in the physically transparent form
\begin{equation}
\begin{aligned}
H_{\rm FW}^{(+)} ={}& m + \frac{\mathbf{p}^2}{2m} - \frac{\mathbf{p}^4}{8m^3} + m v \\
&+ \frac{1}{4m}\left\{k,\,\mathbf{p}^2\right\} + \frac{1}{4m}\,\boldsymbol{\sigma}\cdot\left[\boldsymbol{\nabla}\chi\times\mathbf{p}\right] + \frac{1}{8m}\nabla^2\chi \\
&+ \frac{1}{8m^2}\left\{l,\,\mathbf{p}^4\right\} + \frac{1}{8m^2}\,\boldsymbol{\sigma}\cdot\left[\boldsymbol{\nabla}\eta\times\{\mathbf{p},\,\mathbf{p}^2\}\right] \\
&+ \frac{1}{16m^2}\left\{\nabla^2\eta,\,\mathbf{p}^2\right\} + \mathcal{O}(1/m^3),
\end{aligned}
\label{C27}
\end{equation}
where the additional functions appearing at order \(1/m^2\) are defined as
\begin{align}
l &\equiv \frac{V}{W^4} - 1 - 4f, \label{C28} \\
\eta &\equiv 4f - 2v - \frac{V}{W^2}. \label{C29}
\end{align}
For completeness, we now derive explicitly each sector.

\subsubsection*{Gravitational potential energy}

From Eq.~(\ref{C19}),
\begin{equation}
H_{\rm pot} = m v = m[V- 1].
\label{C30}
\end{equation}

\subsubsection*{Gravitational kinetic correction (order \(1/m\))}

The kinetic correction at order \(1/m\) is given by
\begin{equation}
H_{\rm kin}^{(1/m)} = \frac{1}{4m}\left\{k,\,\mathbf{p}^2\right\},
\label{C31}
\end{equation}
where
\begin{equation}
k = \frac{V}{W^2} - 1 = F^2 - 1.
\label{C32}
\end{equation}

\subsubsection*{Spin-orbit interaction (order \(1/m\))}

The spin-orbit term arises from the commutator \([\mathcal{O},\mathcal{E}]\) and the spin part of \(\mathcal{O}^2\). After careful evaluation, we find
\begin{equation}
H_{\rm SO} = \frac{1}{4m}\,\boldsymbol{\sigma}\cdot\left[\boldsymbol{\nabla}\chi\times\mathbf{p}\right],
\label{C33}
\end{equation}
where
\begin{equation}
\chi = 2f - v= 2\left(\frac{V}{W} - 1\right) - (V - 1) = \frac{2V}{W} - V - 1.
\label{C34}
\end{equation}

\subsubsection*{Darwin term (order \(1/m\))}

The Darwin term comes from the non-spin part of the commutator and from the symmetrisation of derivatives:
\begin{equation}
H_{\rm Darwin} = \frac{1}{8m}\nabla^2\chi.
\label{C35}
\end{equation}

\subsubsection*{Order \(1/m^2\) corrections}

At order \(1/m^2\), we obtain three types of corrections:

\begin{enumerate}
\item \textbf{Modification of the \(\mathbf{p}^4\) term:}
\begin{equation}
H_{\mathbf{p}^4}^{(1/m^2)} = \frac{1}{8m^2}\left\{l,\,\mathbf{p}^4\right\},
\label{C36}
\end{equation}

\item \textbf{Higher-order spin-orbit coupling:}
\begin{equation}
H_{\rm SO}^{(1/m^2)} = \frac{1}{8m^2}\,\boldsymbol{\sigma}\cdot\left[\boldsymbol{\nabla}\eta\times\{\mathbf{p},\,\mathbf{p}^2\}\right],
\label{C37}
\end{equation}

\item \textbf{Non-relativistic Darwin-type correction:}
\begin{equation}
H_{\rm Darwin}^{(1/m^2)} = \frac{1}{16m^2}\left\{\nabla^2\eta,\,\mathbf{p}^2\right\}.
\label{C38}
\end{equation}
\end{enumerate}

\subsection{Final compact form and limiting cases}

For convenience, we may write the Hamiltonian (\ref{C27}) in a more compact form by separating the free part and the gravitational corrections:
\begin{equation}
H_{\rm FW}^{(+)} = H_{\rm free} + H_{\rm grav},
\label{C39}
\end{equation}
where
\begin{equation}
H_{\rm free} = m + \frac{\mathbf{p}^2}{2m} - \frac{\mathbf{p}^4}{8m^3}
\label{C40}
\end{equation}
is the free-particle Hamiltonian, and the gravitational part is
\begin{equation}
\begin{aligned}
H_{\rm grav} ={}& m v + \frac{1}{4m}\left\{k,\,\mathbf{p}^2\right\} + \frac{1}{4m}\,\boldsymbol{\sigma}\cdot\left[\boldsymbol{\nabla}\chi\times\mathbf{p}\right] + \frac{1}{8m}\nabla^2\chi \\
&+ \frac{1}{8m^2}\left\{l,\,\mathbf{p}^4\right\} + \frac{1}{8m^2}\,\boldsymbol{\sigma}\cdot\left[\boldsymbol{\nabla}\eta\times\{\mathbf{p},\,\mathbf{p}^2\}\right] \\
&+ \frac{1}{16m^2}\left\{\nabla^2\eta,\,\mathbf{p}^2\right\} + \mathcal{O}(1/m^3).
\end{aligned}
\label{C41}
\end{equation}
It is important to note here that we have explicitly written out the Hamiltonian up to terms of order $1/m^2$; however, for the analysis in this work, we require only terms of order $1/m$.
\subsection{Others backgrounds}

The general Hamiltonian (\ref{C41}) can now be particularised to any static spherically symmetric metric by substituting the corresponding \(V\) and \(W\). Two important cases are:

\begin{enumerate}
\item \textbf{Schwarzschild geometry:} Taking \(V = W^{-1} = \sqrt{1 - 2M/\rho}\) and expanding to the appropriate order reproduces the well-known post-Newtonian Hamiltonian for a spin-1/2 particle in a Schwarzschild field.

\item \textbf{Einstein--Gauss--Bonnet geometry:} Substituting the EGB expressions from Appendix~\ref{app:EGB_isotropic},
\begin{align}
V(\rho) &= 1 - \frac{M}{\rho} + \frac{2\xi M^2}{\rho^4} + \mathcal{O}(M^2, \xi^2 M^3), \label{C47} \\
W(\rho) &= 1 + \frac{M}{\rho} - \frac{\xi M^2}{2\rho^4} + \mathcal{O}(M^2, \xi^2 M^3), \label{C48}
\end{align}
and keeping terms linear in \(M\) and \(\xi M^2\), we recover the EGB Hamiltonian used in the main text:
\begin{equation}
H_{\mathrm{EGB}}=\frac{2 m \xi M^2}{\rho^4}+\frac{1}{4 m}\left\{\frac{3 \xi M^2}{\rho^4}, \mathbf{p}^2\right\}-\frac{3 \xi M^2}{m \rho^6} \boldsymbol{\sigma} \cdot \mathbf{L}+\frac{9 \xi M^2}{m^2 \rho^6}.
\label{C49}
\end{equation}
\end{enumerate}

\subsection{Remarks on the perturbative expansion}

Several important observations are in order:

\begin{enumerate}
\item \textbf{Generality:} The Hamiltonian (\ref{C41}) is valid for \emph{any} static spherically symmetric metric of the form (\ref{C1}), provided the functions \(V\) and \(W\) are sufficiently smooth. This makes the appendix a self-contained reference for future applications.

\item \textbf{Order of expansion:} We have retained all terms up to order \(1/m^2\) in the FW expansion. This includes the standard \(\mathbf{p}^4\) correction, the gravitational modifications to the kinetic energy, the spin-orbit coupling, the Darwin term, and the higher-order relativistic corrections. Terms of order \(1/m^3\) and higher are consistently neglected.

\item \textbf{Post-Newtonian vs. weak-field:} The expansion in \(1/m\) is independent of the expansion in the gravitational parameters (such as \(M/\rho\) or \(\xi M^2/\rho^4\)). In the main text, we further assume a weak gravitational field and keep only the leading terms in each gravitational parameter. If one desires a fully post-Newtonian Hamiltonian, one would need to retain higher powers of \(M/\rho\) in \(V\) and \(W\).

\item \textbf{Hermiticity:} All operator products have been symmetrised to ensure Hermiticity. The anticommutators \(\{\,,\}\) guarantee that the Hamiltonian is self-adjoint with respect to the flat measure \(d^3x\) \cite{maciel2025gravitational}.

\item \textbf{Spin algebra:} The spin-dependent terms have been evaluated using the standard identities \(\alpha^i\alpha^j = \delta^{ij} + i\epsilon^{ijk}\Sigma^k\) and \(\Sigma^k = \text{diag}(\sigma^k,\sigma^k)\). The final expressions are written in terms of the Pauli matrices \(\boldsymbol{\sigma}\) acting on the positive-energy spinors.
\end{enumerate}

This completes the general derivation of the Foldy--Wouthuysen Hamiltonian for a static spherically symmetric metric in isotropic coordinates.

\bibliographystyle{utphys}
\bibliography{refYBE}

\providecommand{\href}[2]{#2}\begingroup\raggedright\begin{thebibliography}{10}

\bibitem{carroll2019spacetime}
S.~M. Carroll, {\em Spacetime and Geometry: An Introduction to General Relativity}, Cambridge University Press, Cambridge (2019),
\href{https://doi.org/10.1017/9781108770385} {{\tt doi:10.1017/9781108770385}}.

\bibitem{robert1984general}
R.~M. Wald, {\em General Relativity}, University of Chicago Press, Chicago (1984),
\href{https://doi.org/10.7208/chicago/9780226870373.001.0001} {{\tt doi:10.7208/chicago/9780226870373.001.0001}}.

\bibitem{weinberg1973gravitation}
S.~Weinberg, {\em Gravitation and Cosmology: Principles and Applications of the General Theory of Relativity},
John Wiley \& Sons, New York (1972).

\bibitem{Lovelock:1971einstein}
D.~Lovelock, ``{The Einstein tensor and its generalizations},''\href{https://doi.org/10.1063/1.1665613}{{\em J.Math.Phys.}{\bf12} (1971) 498--501}, \href{https://doi.org/10.1063/1.1665613}{{\tt doi:10.1063/1.1665613}}.

\bibitem{Boulware1985}
D.~G. Boulwareand S.~Deser, ``{String-generated gravity models},''\href{https://doi.org/10.1103/PhysRevLett.55.2656}{{\em Phys. Rev. Lett.} {\bf55} (1985) 2656--2660}, \href{https://doi.org/10.1103/PhysRevLett.55.2656}{{\tt doi:10.1103/Phys. Rev. Lett. 55. 2656}}.

\bibitem{glavan2020einstein}
D.~GlavanandC.~Lin, ``{Einstein-Gauss-Bonnet gravity in four-dimensional spacetime},'' \href{https://doi.org/10.1103/PhysRevLett.124.081301}{{\em Phys. Rev. Lett.} {\bf124} (2020) 081301},
\href{https://doi.org/10.1103/PhysRevLett.124.081301}{{\tt doi:10.1103/PhysRevLett.124.081301}}.

\bibitem{fernandes2020derivation}
P.~G.~S. Fernandes, P.~Carrilho,T.~Clifton and D.~J. Mulryne, ``{Derivation of regularized field equations for the Einstein-Gauss-Bonnet theory in four dimensions},'' \href{https://doi.org/10.1103/PhysRevD.102.024025}{{\em Phys. Rev. D}{ \bf102} (2020) 024025,}
\href{https://doi.org/10.1103/PhysRevD.102.024025}{{\tt doi:10.1103/PhysRevD.102.024025}},

\bibitem{hennigar2020lower}
R.~A. Hennigar, D.~Kubiz\v{n}\'ak, R.~B. Mann and C.~Pollack, ``{Lower-dimensional Gauss--Bonnet gravity and BTZ black holes},''\href{https://doi.org/10.1016/j.physletb.2020.135657} {{\em Phys. Lett. B} {\bf 808} (2020) 135657},
\href{https://doi.org/10.1016/j.physletb.2020.135657}{{\tt doi:10.1016/j.physletb.2020.135657}},

\bibitem{lu2020horndeski}
H.~L\"uand Y.~Pang, ``{Horndeski gravity as $D\to4$ limit of Gauss--Bonnet},'' \href{https://doi.org/10.1016/j.physletb.2020.135717}{{\em Phys. Lett. B}{\bf 809} (2020) 135717},
\href{https://doi.org/10.1016/j.physletb.2020.135717}{{\tt doi:10.1016/j.physletb.2020.135717}}.

\bibitem{aoki2020consistent}
K.~Aoki,M.~A.GorjiandS.~Mukohyama, ``{A consistent theory of$D\to4$ Einstein--Gauss--Bonnet gravity},'' \href{https://doi.org/10.1016/j.physletb.2020.135843} {{\em Phys.Lett.B} {\bf810} (2020) 135843},
\href{https://doi.org/10.1016/j.physletb.2020.135843}{{\tt doi:10.1016/j.physletb.2020.135843}}.

\bibitem{fernandes2022Review}
P.~G.~S.Fernandes, P.~Carrilho, T.~Cliftonand D.~J.Mulryne, ``{The 4D-Einstein-Gauss-Bonnet theory of gravity:A review},'' \href{https://doi.org/10.1088/1361-6382/ac500a}{{\em Class. Quantum Grav.} {\bf39} (2022) 063001},
\href{https://doi.org/10.1088/1361-6382/ac500a}{{\tt doi:10.1088/1361-6382/ac500a}}.

\bibitem{cvetivc2002black}
M.~Cveti\v{c}, S.~Nojiriand S.~D. Odintsov,``{Black hole thermodynamics and negative entropy in de Sitter and anti-de Sitter Einstein-Gauss-Bonnet gravity},'' \href{https://doi.org/10.1016/S0550-3213(02)00075-5}{{\em Nucl. Phys. B} {\bf628} (2002) 295-330},
\href{https://doi.org/10.1016/S0550-3213(02)00075-5}{{\tt doi:10.1016/S0550-3213(02)00075-5}}.

\bibitem{nojiri2024propagation}
S.~Nojiri, S.~D.O dintsovand V.~K. Oikonomou, ``{Propagation of gravitational waves in Einstein--Gauss--Bonnet gravity for cosmological and spherically symmetric spacetimes},'' \href{https://doi.org/10.1103/PhysRevD.109.044046}{{\em Phys. Rev. D} {\bf109} (2024) 044046},
\href{https://doi.org/10.1103/PhysRevD.109.044046}{{\tt doi:10.1103/PhysRevD.109.044046}}.

\bibitem{nojiri2024}
S.~NojiriandS.~D. Odintsov, ``{Propagation speed of gravitational wave in scalar--Einstein--Gauss--Bonnet gravity},'' \href{https://doi.org/10.1016/j.nuclphysb.2023.116423}{{\em Nucl. Phys. B} {\bf998} (2024) 116423},
\href{https://doi.org/10.1016/j.nuclphysb.2023.116423}{{\tt doi:10.1016/j.nuclphysb.2023.116423}}.

\bibitem{nojiri2017modified}
S.~Nojiri, S.~D. Odintsov and V.~K.Oikonomou, ``{Modified gravity theories on a nutshell:Inflation, bounce and late-time evolution},'' \href{https://doi.org/10.1016/j.physrep.2017.06.001}{{\em Phys. Rept.} {\bf692} (2017) 1--104},
\href{https://doi.org/10.1016/j.physrep.2017.06.001}{{\tt doi:10.1016/j.physrep.2017.06.001}}.

\bibitem{churilova2021quasinormal}
M.~S. Churilova, ``{Quasinormal modes of the Dirac field in the consistent 4D Einstein--Gauss--Bonnet gravity},''\href{https://doi.org/10.1016/j.dark.2020.100748}{{\em Phys. Dark Univ.} {\bf 31} (2021) 100748},
\href{https://doi.org/10.1016/j.dark.2020.100748} {{\tt doi:10.1016/j.dark.2020.100748}}.

\bibitem{churilova2021quasinormal2}
M.~S. Churilova, ``{Quasinormal modes of the test fields in the consistent 4D Einstein--Gauss--Bonnet--(anti)de Sitter gravity},''\href{https://doi.org/10.1016/j.aop.2021.168425}{{\em Ann. Phys.} {\bf 427} (2021) 168425},
\href{https://doi.org/10.1016/j.aop.2021.168425}{{\tt doi:10.1016/j.aop.2021.168425}}.

\bibitem{churilova2021wormholes}
M.~S. Churilova, R.~A. Konoplya, Z.~Stuchlík and A.~Zhidenko, ``{Wormholes without exotic matter: Quasinormal modes, echoes and shadows},''\href{https://doi.org/10.1088/1475-7516/2021/10/010}{{\em JCAP} {\bf 10} (2021) 010},
\href{https://doi.org/10.1088/1475-7516/2021/10/010}{{\tt doi:10.1088/1475-7516/2021/10/010}}.

\bibitem{clifton2020observational}
T.~Clifton, P.~Carrilho, P.~G.~S. Fernandes and D.~J. Mulryne, ``{Observational constraints on the regularized 4D Einstein--Gauss--Bonnet theory of gravity},''\href{https://doi.org/10.1103/PhysRevD.102.084005}{{\em Phys. Rev. D} {\bf 102} (2020) 084005}, \href{https://doi.org/10.1103/PhysRevD.102.084005}{{\tt doi:10.1103/PhysRevD.102.084005}}.

\bibitem{charmousis2022astrophysical}
C.~Charmousis, A.~Leh\'ebel, E.~Smyrniotis and N.~Stergioulas, ``{Astrophysical constraints on compact objects in 4D Einstein--Gauss--Bonnet gravity},''\href{https://doi.org/10.1088/1475-7516/2022/02/033}{{\em JCAP} {\bf 02} (2022) 033},\href{https://doi.org/10.1088/1475-7516/2022/02/033}{{\tt doi:10.1088/1475-7516/2022/02/033}},

\bibitem{brill1957interaction}
D.~R. Brill and J.~A. Wheeler, ``{Interaction of neutrinos and gravitational fields},''\href{https://doi.org/10.1103/RevModPhys.29.465}{{\em Rev. Mod. Phys.} {\bf 29} (1957) 465--479},
\href{https://doi.org/10.1103/RevModPhys.29.465}{{\tt doi:10.1103/RevModPhys.29.465}}.

\bibitem{parker2009quantum}
L.~Parker and D.~Toms, ``{{\em Quantum Field Theory in Curved Spacetime: Quantized Fields and Gravity}}'',
Cambridge University Press, Cambridge (2009), \href{https://doi.org/10.1017/CBO9780511813924}{{\tt doi:10.1017/CBO9780511813924}}.

\bibitem{obukhov2013spin}
Y.~N. Obukhov, A.~J. Silenko and O.~V. Teryaev, ``{Spin in an arbitrary gravitational field},'' \href{https://doi.org/10.1103/PhysRevD.88.084014}{{\em Phys. Rev. D} {\bf 88} (2013) 084014},
\href{https://doi.org/10.1103/PhysRevD.88.084014}{{\tt doi:10.1103/PhysRevD.88.084014}}.

\bibitem{foldy1950dirac}
L.~L. Foldy and S.~A. Wouthuysen, ``{On the Dirac theory of spin $1/2$ particles and its non-relativistic limit},''\href{https://doi.org/10.1103/PhysRev.78.29}{{\em Phys. Rev.} {\bf 78} (1950) 29--36},
\href{https://doi.org/10.1103/PhysRev.78.29}{{\tt doi:10.1103/PhysRev.78.29}}.

\bibitem{obukhov2001spin}
Y.~N. Obukhov, ``{Spin, gravity, and inertia},'' \href{https://doi.org/10.1103/PhysRevLett.86.192}{{\em Phys. Rev. Lett.} {\bf 86} (2001) 192--195},
\href{https://doi.org/10.1103/PhysRevLett.86.192}{{\tt doi:10.1103/PhysRevLett.86.192}}.

\bibitem{jentschura2013nonrelativistic}
U.~D. Jentschura and J.~H. Noble, ``{Nonrelativistic limit of the Dirac--Schwarzschild Hamiltonian: Gravitational Zitterbewegung and gravitational spin-orbit coupling},'' \href{https://doi.org/10.1103/PhysRevA.88.022121}{{\em Phys. Rev. A} {\bf 88} (2013) 022121},
\href{https://doi.org/10.1103/PhysRevA.88.022121}{{\tt doi:10.1103/PhysRevA.88.022121}}.

\bibitem{MACIEL2026170696}
E.~Maciel, ``{Dynamics for spin-1/2 particles in Einstein--Gauss--Bonnet gravity},''\href{https://doi.org/10.1016/j.aop.2026.170696} {{\em Ann. Phys.} (2026) 170696},
\href{https://doi.org/10.1016/j.aop.2026.170696}{{\tt doi:10.1016/j.aop.2026.170696}}.

\bibitem{cantrell2010inclination}
A.~G. Cantrell, C.~D. Bailyn, J.~A. Orosz, J.~E. McClintock, R.~A. Remillard, C.~S. Froning, J.~Neilsen, D.~M. Gelino and L.~Gou, ``{The inclination of the soft X-ray transient A0620--00 and the mass of its black hole},'' \href{https://doi.org/10.1088/0004-637X/710/2/1127} {{\em Astrophys. J.} {\bf 710} (2010) 1127--1141},
\href{https://doi.org/10.1088/0004-637X/710/2/1127}{{\tt doi:10.1088/0004-637X/710/2/1127}}.

\bibitem{gou2010spin}
L.~Gou, J.~E. McClintock, J.~F. Steiner, R.~Narayan, A.~G. Cantrell, C.~D. Bailyn and J.~A. Orosz, ``{The spin of the black hole in the soft X-ray transient A0620--00},'' \href{https://doi.org/10.1088/2041-8205/718/2/L122}{{\em Astrophys. J. Lett.} {\bf 718} (2010) L122--L126},
\href{https://doi.org/10.1088/2041-8205/718/2/L122} {{\tt doi:10.1088/2041-8205/718/2/L122}}.





\bibitem{zumino1986gravity}
B.~Zumino, ``{Gravity theories in more than four dimensions},'' \href{https://doi.org/10.1016/0370-1573(86)90076-1}{{\em Phys. Rept.}{\bf137}(1986)109--114},
\href{https://doi.org/10.1016/0370-1573(86)90076-1}{{\tt doi:10.1016/0370-1573(86)90076-1}}.

\bibitem{gurses2020there}
M.~G\"urses, T.~C. \c{S}i\c{s}manandB.~Tekin, ``{Is the reanovel Einstein--Gauss--Bonnet theory in fourdimensions?},'' \href{https://doi.org/10.1140/epjc/s10052-020-8200-7}{{\em Eur. Phys. J. C} {\bf80} (2020) 647}, \href{https://doi.org/10.1140/epjc/s10052-020-8200-7}{{\tt doi:10.1140/epjc/s10052-020-8200-7}}.

\bibitem{wheeler1986symmetric}
J.~T. Wheeler, ``{Symmetric solutions to the maximally Gauss--Bonnet extended Einstein equations},'' \href{https://doi.org/10.1016/0550-3213(86)90388-3}{{\em Nucl. Phys. B} {\bf 273} (1986) 732--748},
\href{https://doi.org/10.1016/0550-3213(86)90388-3}{{\tt doi:10.1016/0550-3213(86)90388-3}}.

\bibitem{hennigar2020taking}
R.~A. Hennigar, D.~Kubiz\v{n}\'ak, R.~B. MannandC.~Pollack. ``{On taking the $D\to4$ limit of Gauss--Bonnet gravity: Theory and solutions}," \href{https://doi.org/10.1007/JHEP07(2020)027}{{\em JHEP} {\bf07} (2020) 027},
\href{https://doi.org/10.1007/JHEP07(2020)027}{{\tt doi:10.1007/JHEP07(2020)027}}.

\bibitem{fernandes2020charged}
P.~G.~S.Fernandes, ``{Charged black holes in AdS spaces in 4D Einstein-Gauss-Bonnet gravity},'' \href{https://doi.org/10.1016/j.physletb.2020.135468}  {{\em Phys. Lett. B} {\bf805} (2020), 135468},
\href{https://doi.org/10.1016/j.physletb.2020.135468}{{\tt doi:10.1016/j.physletb.2020.135468}}.

\bibitem{obukhov2009spin}
Y.~N. Obukhov, A.~J. Silenko and O.~V. Teryaev, ``{Spin dynamics in gravitational fields of rotating bodies and the equivalence principle},'' \href{https://doi.org/10.1103/PhysRevD.80.064044} {{\em Phys. Rev. D} {\bf 80} (2009) 064044},
\href{https://doi.org/10.1103/PhysRevD.80.064044}{{\tt doi:10.1103/PhysRevD.80.064044}}.

\bibitem{silenko2008foldy}
A.~J. Silenko, ``{Foldy--Wouthuysen transformation and semiclassical limit for relativistic particles in strong external fields},'' \href{https://doi.org/10.1103/PhysRevA.77.012116}{{\em Phys. Rev. A} {\bf 77} (2008) 012116},
\href{https://doi.org/10.1103/PhysRevA.77.012116}{{\tt doi:10.1103/PhysRevA.77.012116}}.

\bibitem{silenko2005semiclassical}
A.~J. Silenko and O.~V. Teryaev, ``{Semiclassical limit for Dirac particles interacting with a gravitational field},'' \href{https://doi.org/10.1103/PhysRevD.71.064016}{{\em Phys. Rev. D} {\bf 71} (2005) 064016},
\href{https://doi.org/10.1103/PhysRevD.71.064016}{{\tt doi:10.1103/PhysRevD.71.064016}}.

\bibitem{obukhov2011dirac}
Y.~N. Obukhov, A.~J. Silenko and O.~V. Teryaev, ``{Dirac fermions in strong gravitational fields},'' \href{https://doi.org/10.1103/PhysRevD.84.024025}{{\em Phys. Rev. D} {\bf 84} (2011) 024025},
\href{https://doi.org/10.1103/PhysRevD.84.024025}{{\tt doi:10.1103/PhysRevD.84.024025}},

\bibitem{guo2020innermost}
M.~GuoandP.-C.Li, ``{Inner most stable circular orbit and shadow of the 4D Einstein--Gauss--Bonnet black hole},''\href{https://doi.org/10.1140/epjc/s10052-020-8164-7}{{\em Eur. Phys. J. C} {\bf80} (2020) 588},
\href{https://doi.org/10.1140/epjc/s10052-020-8164-7}{{\tt doi:10.1140/epjc/s10052-020-8164-7}}.

\bibitem{li2021tidal}
J.~Li,S.~ChenandJ.~Jing, ``{Tidal effects in 4D Einstein--Gauss--Bonnet black hole spacetime},'' \href{https://doi.org/10.1140/epjc/s10052-021-09372-w}{{\em Eur. Phys. J. C} {\bf81} (2021) 590},
\href{https://doi.org/10.1140/epjc/s10052-021-09372-w}{{\tt doi:10.1140/epjc/s10052-021-09372-w}}.

\bibitem{vieira2023quasibound}
H.~S. Vieira, ``{Quasibound states of analytic black-hole configurations in three and four dimensions},''\href{https://doi.org/10.1103/PhysRevD.107.104011}{{\em Phys. Rev. D} {\bf 107} (2023) 104011},
\href{https://doi.org/10.1103/PhysRevD.107.104011}{{\tt doi:10.1103/PhysRevD.107.104011}}.

\bibitem{davies1980quantum}
P.~C.~W. Davies, ``{Quantum fields in curved space},'' in {{\em General Relativity and Gravitation}}, Vol.~1,
ed. A.~Held, Plenum Press, New York (1980), p.~255.

\bibitem{collas2019dirac}
P.~Collas and D.~Klein, ``{{\em The Dirac Equation in Curved Spacetime: A Guide for Calculations}}',
Springer, Cham (2019), \href{https://doi.org/10.1007/978-3-030-14825-6}{{\tt doi:10.1007/978-3-030-14825-6}}.

\bibitem{maciel2025gravitational}
E.~B. Maciel, M.~A. Anacleto and E.~Passos, ``{Gravitational dipole moment in braneworld model},'' \href{https://doi.org/10.1140/epjp/s13360-025-06605-5}{{\em Eur. Phys. J. Plus} {\bf 140} (2025) 678},
\href{https://doi.org/10.1140/epjp/s13360-025-06605-5}{{\tt doi:10.1140/epjp/s13360-025-06605-5}}.

\bibitem{eriksen1958foldy}
E.~Eriksen,``{Foldy--Wouthuysen transformation. Exact solution with generalization to the two-particle problem},''\href{https://doi.org/10.1103/PhysRev.111.1011}{{\em Phys. Rev.} {\bf 111} (1958) 1011--1016},
\href{https://doi.org/10.1103/PhysRev.111.1011}{{\tt doi:10.1103/PhysRev.111.1011}}.

\bibitem{silenko2015general}
A.~J. Silenko, ``{General dynamics of tensor polarization of particles and nuclei in external fields},''\href{https://doi.org/10.1088/0954-3899/42/7/075109}{{\em J. Phys. G: Nucl. Part. Phys.} {\bf 42} (2015) 075109},
\href{https://doi.org/10.1088/0954-3899/42/7/075109}{{\tt doi:10.1088/0954-3899/42/7/075109}}.

\bibitem{bjorken1965relativistic}
J.~D. Bjorken and S.~D. Drell, ``{Relativistic Quantum Mechanics}', {{\em McGraw--Hill, New York (1964)}}.

\bibitem{greiner1990relativistic}
W.~Greiner, ``{Relativistic Quantum Mechanics: Wave Equations}'' {{\em Springer, Berlin, Heidelberg (1990)}},,
\href{https://doi.org/10.1007/978-3-662-02634-2}{{\tt doi:10.1007/978-3-662-02634-2}}.

\bibitem{liboff2003introductory}
R.~L. Liboff,{{\em Introductory Quantum Mechanics}}, 4th ed., Addison--Wesley, San Francisco (2003).

\bibitem{merzbacher1998quantum}
E.~Merzbacher, {{\em Quantum Mechanics}}, 3rd ed., John Wiley \& Sons, New York (1998).

\bibitem{landau1958quantum}
L.~D. Landau and E.~M. Lifshitz,{{\em Quantum Mechanics: Non-Relativistic Theory}}, Course of Theoretical Physics, Vol.~3, Pergamon Press, London (1958).

\bibitem{shen2005spin}
S.-Q.~Shen, ``{Spin transverse force on spin current in an electric field},''\href{https://doi.org/10.1103/PhysRevLett.95.187203}{{\em Phys. Rev. Lett.} {\bf 95} (2005) 187203},
\href{https://doi.org/10.1103/PhysRevLett.95.187203}{{\tt doi:10.1103/PhysRevLett.95.187203}},













\end{thebibliography}\endgroup

\end{document}